\documentclass[aps, prd, amsmath, floats, floatfix, twocolumn,superscriptaddress, nofootinbib, showpacs,reprint]{revtex4}
\usepackage{grffile}
\usepackage{graphicx}
\usepackage{amsmath,amssymb}
\usepackage{amsfonts}
\usepackage{xspace} 
\usepackage[usenames]{color}
\usepackage{dcolumn}
\usepackage{bm}
\usepackage{mathrsfs}

\usepackage{ulem}
\usepackage{appendix}

\definecolor {darkgreen}{rgb}{0.2,0.7,0.2}

\newcommand{\eq}{\begin{equation}}
\newcommand{\be}{\begin{equation}}
\newcommand{\eeq}{\end{equation}}
\newcommand{\ee}{\end{equation}}

\newcommand{\apjl}{Astrophys.\ J.\ Lett.\ }

\begin{document}

\title{Dynamical scalarization of neutron stars in  scalar-tensor gravity theories}

\author{Carlos Palenzuela}
\affiliation{Canadian Institute for Theoretical Astrophysics, Toronto, Ontario M5S 3H8, Canada}
\author{Enrico Barausse}
\affiliation{Institut d'Astrophysique de Paris/CNRS, 98bis boulevard Arago, 75014 Paris, France}
\author{Marcelo Ponce} 
\affiliation{Department of Physics, University of Guelph, Guelph, Ontario N1G 2W1, Canada}
\author{Luis Lehner}
\affiliation{Perimeter Institute for Theoretical Physics, Waterloo, Ontario N2L 2Y5, Canada}
\affiliation{CIFAR, Cosmology \& Gravity Program, Canada}

\begin{abstract}

We present a framework to study generic neutron-star binaries in scalar-tensor theories of gravity.
Our formalism achieves this goal by suitably interfacing a post-Newtonian orbital evolution
(described by a set of  ordinary differential equations) with 
a set of non-linear algebraic equations, which provide
a description of the scalar charge of each binary's component
along the evolution in terms of isolated-star data.
We validate this semi-analytical procedure by comparing its results to those of 
fully general-relativistic simulations, and use it to investigate the behavior of
binary systems in large portions of the parameter space of scalar-tensor theories.
This allows us to shed further light on the phenomenon of ``dynamical scalarization'',
which we uncovered in [Barausse, Palenzuela, Ponce and Lehner, Phys. Rev. D 87,  081506(R) (2013)]
and which takes place in tight binaries, even for stars that have exactly zero scalar charge in isolation. 
We also employ our formalism to study representative binary systems,  obtain their gravitational-wave signals
and discuss the extent to which deviations from General Relativity can be detected. The insights gained by
this framework allow us to additionally  
show that eccentric binaries can undergo scalarization/de-scalarization phenomena.
\end{abstract}

\date{\today \hspace{0.2truecm}}

\pacs{04.25.-g,04.25.D-,04.30.-w}

\maketitle

\section{Introduction}

Scalar-tensor theories of gravity~\cite{wagoner,bergmann,nordtvedt,fierz,jordan,BD} are among the oldest and most natural alternatives to General Relativity (GR). In addition to
the usual spin-2 graviton of GR, these theories present an additional spin-0 graviton polarization, i.e.
gravity is encoded not only by the metric, but also by a gravitational scalar field. Possible 
candidates for this scalar field include, for instance, the dilaton field of string theory.
 This scalar field couples non-minimally
to the metric, but does not couple directly to matter
(because if it did, it would introduce a fifth force in the interactions of matter, which 
is not observed experimentally). Nevertheless, because the scalar couples non-minimally to the metric, which in turn
is coupled to matter through gravity, an effective coupling appears between the scalar field and matter. Being mediated
by gravity, however, this coupling is generally weak and only important in suitable
 astrophysical contexts. This effective coupling
depends, for generic scalar tensor theories, on the local value of the scalar field, i.e. the effective (dimensionless) coupling constant $\tilde{\alpha}( \tilde{\varphi} )$
can be Taylor-expanded around the Minkowski vacuum $\tilde{\varphi}_0=$ const of the (dimensionless) scalar field $\tilde{\varphi}$ as~\cite{damour_farese_long,spontaneous_scalarization,spontaneous_scalarization_bis}
\begin{equation}
\tilde{\alpha}( \tilde{\varphi} )\approx \frac{1}{\sqrt{3+2 \omega_0}}-\tilde{\beta} (\tilde{\varphi}-\tilde{\varphi}_0) + {\cal O} (\tilde{\varphi}-\tilde{\varphi}_0)^2\,,\label{coupling}
\end{equation}
where $\omega_0$ and $\tilde{\beta}$ are dimensionless constants.

The simplest theory in the class of scalar-tensor theories was proposed in the 1950's by Fierz~\cite{fierz} and 
Jordan~\cite{jordan}, and later rediscovered
by Brans and Dicke~\cite{BD}, hence it is usually called Fierz-Jordan-Brans-Dicke (FJBD) theory. 
This theory truncates the effective coupling \eqref{coupling}
at the lowest order, i.e. it assumes $\tilde{\beta}=0$ and neglects the higher-order $ {\cal O} (\tilde{\varphi}-\tilde{\varphi}_0)^2$ terms.
Because gravity behaves very much as predicted by GR in the solar system, the 
effective coupling of FJBD theory is then constrained to very small values by solar-system tests, and in particular the Cassini mission~\cite{solar_system}, 
which requires $\omega_0>40000$. Under such a tight experimental constraint, FJBD theory is essentially indistinguishable from GR
as far as astrophysical tests are concerned (except perhaps with space-based gravitational-wave detectors such as eLISA~\cite{elisa,berti_buonanno_will,floating_orbits,gw_ST1}).

It is important to stress that solar-system experiments are characterized by velocities $v \lesssim 2 \times 10^{-4} c$, and weak gravitational fields $\Phi_{\rm Newt}/c^2\lesssim 10^{-6}$. Thus
they do not constrain the strong-field, very relativistic regime where the most surprising predictions of 
GR, such as black holes (BHs) and neutron stars (NS), arise.
In this regime, the FJBD assumption of retaining only the first term in the expansion \eqref{coupling}
is not necessarily justified from an effective field theory point of view. 
Furthermore, the second term in that expansion has been shown 
to produce dramatic effects in strong-field regimes such as 
the interior structure of NSs~\cite{spontaneous_scalarization,spontaneous_scalarization_bis} (see also Refs.~\cite{sotani,kokkotas} for the rotating NS case).
More specifically, if $\tilde{\beta}$ is sufficiently negative and for NS compactnesses above a critical value, the trivial vacuum
of the scalar field becomes unstable.
It then becomes energetically favorable for the scalar field to settle on a non trivial configuration 
inside the NS (``spontaneous scalarization''). This new non-trivial vacuum
not only affects the relation between the NS mass and its radius, but also the orbital evolution of binary NS systems.
Such behavior arises because scalarization
enhances the gravitational attraction between the binary's components, and triggers the emission of dipolar
scalar radiation~\cite{eardley_dipole,damour_farese_long,will_zaglauer}. These effects allowed constraints 
to be placed on the constant $\tilde{\beta}$ using binary pulsar
data~\cite{spontaneous_scalarization_bis,ST_pulsars1,ST_pulsars2,ST_pulsars3}, which require $\tilde{\beta}>-4.5$.
It has also been suggested that negative values of  $\tilde{\beta}$ might produce significant effects in supernova 
explosions and in inflationary scenarios~\cite{supernovae_inflation}, as well as affect the gravitational wave 
spectrum of vibrating scalarized NSs~\cite{harada,NM_ST}.

All these results, however, probe the strong-field \textit{mildly relativistic} regime, because they involve 
either static/stationary configurations or
velocities much smaller than the speed of light (e.g. binary pulsars have velocities $v\sim 10^{-3} c$). 
The first investigations of the strong-field relativistic regime of scalar tensor theories were
performed by Ref.~\cite{GATech} for a close binary system of BHs, and by ourselves in Ref.~\cite{ourselves} for 
close binary NS systems.
Not surprisingly, the behavior of binary BH systems turns out to be essentially indistinguishable from GR~\cite{GATech}, as
expected from results obtained in isolated BHs~\cite{hawking,thorne} and for binary BH systems in the weak-field, 
mildly relativistic regime~\cite{will_zaglauer,damour_farese_long,will_latest}. On the other hand,
the strong effects and deviations discussed in Ref.~\cite{ourselves} were unanticipated from the 
intuition drawn from weak-field, mildly relativistic analysis.
 
In particular, we showed~\cite{ourselves} that the second term in the expansion \eqref{coupling}  can have
strong consequences in the late stages of the evolution of NS binaries, even in cases
where no effects are observed in the  weak-field, mildly relativistic regime. (See also Ref.~\cite{shibata_buonanno} 
for later exploration and further evidence of this result). 
In fact, we observe large deviations away from the GR behavior at separations much smaller than those probed by binary pulsars, 
providing signals that are at least in principle observable with existing gravitational-wave detectors  such as Advanced LIGO/Virgo. 
These facilities are expected to detect (from several to hundreds) NS binaries
per year of operation~\cite{merger_rates}, and are sensitive from tens of Hz up to  a kHz range. While the plunge and merger of binary NSs
is outside Advanced LIGO/Virgo's sensitivity band in GR, our results~\cite{ourselves} show that negative values of $\tilde{\beta}$ 
can trigger earlier plunges in  scalar-tensor theories, thus moving the plunge and merger to frequencies of 
about 500-600 Hz, within the detectors' reach.

In this paper, we provide a deeper insight into the physical origin of these earlier plunges. We 
show that according to the binary system's parameters, they can either be triggered by the \textit{induced 
scalarization} (IS) of one (initially non-scalarized) star under
the influence of the ``external'' scalar field produced by the other (initially scalarized) star, or by a 
genuine phase transition of the full, initially non-scalarized system,  i.e. a \textit{dynamical scalarization} (DS).
While the  interpretation of IS is quite straightforward, the physical process at play in DS
is less obvious. Expanding on our discussion in Ref.~\cite{ourselves}, we show that the DS phase transition generalizes 
the {\it spontaneous scalarization} of Refs.~\cite{spontaneous_scalarization,spontaneous_scalarization_bis}. 
Indeed, spontaneous scalarization 
takes place when an isolated NS has a compactness $G M/(R c^2)$ (where $M$ and $R$ are the stellar mass and radius) above 
a certain critical threshold, while we argue that DS 
occurs when a suitably defined ``effective'' compactness of the binary rises above a certain critical value in the last
stages of the post-Newtonian (PN) inspiral~\footnote{A closely related effect has been recently presented in the context of BH systems
interacting with matter~\cite{vitorthomas}.}. Note that DS thus allows for a richer phenomenology,
enabling scalarization of stars unable to scalarize in isolation; stronger scalarization of compact binaries
as their orbit shrinks;  and even
a {\em transient dynamical scalarization} in (close) eccentric encounters. All these
phenomena produce strong deviations away from the GR behavior, which may be observable with electromagnetic 
or gravitational probes.

To obtain the generic behavior of NS binaries prior to merger in an efficient, semi-analytical form,
we present a formalism that describes the orbital evolution within an improved version of the PN dynamics for scalar-tensor 
theories. More specifically, we modify the PN equations of motion for binary systems~\cite{will_latest} to account for  
 the changes in the stars' scalar charges produced by IS and DS. 
Also, we unify the treatment of 
these two phenomena, and describe their effects by a system of non-linear
algebraic equations, which we solve at each step of the orbital evolution to compute the scalar charges. 
For simplicity, we restrict here to polytropic equations of state, but our method can be straightforwardly
modified to consider more general equations of state. 
This approach, besides highlighting the
physical origin of the deviations from GR, provides an efficient and inexpensive way to compute gravitational-wave templates for
Advanced LIGO/Virgo. These can be used to devise strategies to detect these effects once signals from NS binaries are detected, 
a problem which will be the subject of future work.

Throughout this work we set $c=1$, but reinstate powers of $c$ to keep track of the various PN orders when needed.

\section{Motion in scalar-tensor theories}\label{theory_motion}

Let us consider a scalar-tensor theory with action
\begin{equation}
\label{Jframe_action}
S=\!\!\int d^4 x\frac{\sqrt{-g}}{2\kappa}\left[\phi R-\frac{\omega(\phi)}{\phi} \partial_\mu \phi \partial^\mu \phi
\right]+ S_M[g_{\mu\nu},\psi]\,,
\end{equation}
where  $\kappa=8 \pi G$, 
$R$ and $g$ are respectively the Ricci scalar and the determinant of the metric,
$\phi$ is the gravitational scalar field,  $\omega(\phi)$ is a function that characterizes the theory,
and we denote the degrees of freedom of matter collectively by $\psi$.
As can be seen, the matter fields are not coupled directly to $\phi$ to avoid 
producing a scalar ``fifth force'', as discussed earlier.
We stress that Eq.~\eqref{Jframe_action} is not the most generic action for scalar tensor theories with second-order field equations
(because for instance the scalar field may have a potential, or there may be Galileon-type
terms in the action~\cite{galileon1,galileon2}), but is general enough to highlight the
physics of spontaneous scalarization, DS and IS. 

The action \eqref{Jframe_action} is often referred to as the ``Jordan-frame'' action. An alternative
form for the action (usually called ``Einstein-frame action'') can be obtained by introducing a new metric $g^{E}_{\mu\nu}$, conformally related
to the Jordan frame metric by $g^E_{\mu\nu}=\phi\, g_{\mu\nu}$, and a new scalar field $\varphi$ defined (up to integration constants) by $({{\rm d}\log \phi}/{{\rm d}\varphi})^2={2\kappa}/[{3+2 \omega(\phi)}]$.
In terms of these variables the action \eqref{Jframe_action} becomes 
\begin{equation}
\label{einframe}\!
\!S=\!\!\int\!d^4 x \sqrt{-g^E} \left( \frac{R^E}{2\kappa}\!-\!\frac{1}{2}g_E^{\mu\nu} \partial_\mu\varphi \partial_\nu\varphi
\right) \! +S_M\!\!\left[\frac{g^E_{\mu\nu}}{\phi(\varphi)},\psi\right]\!\!\!
\end{equation}
Note that the matter fields $\psi$ still couple to the physical metric  $g_{\mu\nu}=g^E_{\mu\nu}/\phi$, i.e. weakly gravitating bodies 
follow geodesics of the physical metric and \textit{not} the ones of the Einstein frame metric. The advantage of using Einstein frame
variables is that \text{in vacuum} the metric $g^E_{\mu\nu}$ and the scalar field couple minimally, as can be seen from the action
\eqref{einframe}. This means
 that at the linear level (on a flat background) the Einstein frame metric and the scalar field decouple, 
i.e. the Einstein-frame metric represents the spin-2 graviton polarizations, while the scalar field represents
the spin-0 polarization. The spin-2 and spin-0 polarizations are instead mixed in the Jordan-frame metric. 

Variation of the action \eqref{einframe} yields the following field equations
\begin{gather}\label{einstein}
G_{\mu\nu}^{E}=\kappa \left(T^\varphi_{\mu\nu} + T^E_{\mu\nu} \right),\\
\Box^E \varphi = \frac12 
\frac{{\rm d}\log \phi}{{\rm d}\varphi} T_E\label{KG} \,,\\
\nabla_\mu^E T_E^{\mu\nu}=-\frac{1}{2} T_E 
\frac{{\rm d}\log \phi}{{\rm d}\varphi} 
g_E^{\mu\nu} \partial_\mu \varphi\,,\label{scalarT}
\end{gather}
where indices are raised/lowered with the Einstein-frame metric.
Also, $T_E\equiv T_E^{\mu\nu}g^E_{\mu\nu}$, and the stress energy tensors for matter and the scalar field in the Einstein frame are defined as
\begin{eqnarray} 
T_E^{\mu\nu}&=&\frac{2}{\sqrt{-g^E}}\frac{\delta S_M}{\delta g^E_{\mu\nu}} \,\,\,\,\, \mbox{and} \\
T^\varphi_{\mu\nu} &=&\partial_\mu \varphi \partial_\nu \varphi- \frac{g^E_{\mu\nu}}{2} g_E^{\alpha\beta} \partial_\alpha \varphi
\partial_\beta \varphi\, 
\end{eqnarray} 
while the relation to the ``physical'' matter stress-energy tensor (i.e. the one in the Jordan frame) is given by
$
T_E^{\mu\nu}={T^{\mu\nu}}{\phi^{-3}} \, \, , \, \, T^E_{\mu\nu}={T_{\mu\nu}}{\phi^{-1}}.
$

From Eq.~\eqref{KG}, it is clear that a coupling appears between the Einstein-frame scalar field $\varphi$ (or the Jordan-frame scalar field $\phi$) and matter.
First, this implies that in vacuum (i.e. in the absence of matter) the scalar field is not excited, hence
the vacuum solutions of GR (e.g. Minkowski, BHs, etc.) are still solutions, with $\varphi=\varphi_0=$ const.
Second, because in the original Jordan-frame action \eqref{Jframe_action} the scalar field does not couple to matter directly, 
it is clear that this coupling simply appears because the scalar field couples non-minimally to the Jordan-frame metric, which is in turn 
coupled to matter via gravity (i.e. by the Einstein equations). In fact, Eq.~\eqref{KG} can also be derived from the Jordan-frame action, by combining the
equation of motion for the scalar field with the trace of the Einstein equations. 
Introducing a dimensionless scalar field $\tilde{\varphi}=(4\pi G)^{1/2} \varphi$, we can characterize this gravity-mediated effective coupling
by the dimensionless coupling constant 
\begin{equation}
\tilde{\alpha}\equiv \frac12 
\frac{{\rm d}\log \phi}{{\rm d}\tilde{\varphi}}= \frac{1}{\sqrt{3+2\omega(\varphi)}}\,.
\end{equation}
Comparing this equation to Eq.~\eqref{coupling} we obtain 
\begin{equation}
\phi = \exp[-\beta \varphi^2 + {\cal O}(\varphi-\varphi_0)^3] \,,
\end{equation}
where have defined $\beta= 4\pi G \tilde{\beta}$ and 
\begin{equation}\label{varphi0}
\varphi_0=-\frac{2(G\pi)^{1/2}}{\beta (3+2 \omega_0)^{1/2}}\,
\end{equation}
is to be interpreted as the asymptotic value of the scalar field at spatial infinity.
In what follows, and as done e.g. also in Ref.~\cite{ourselves}, we will therefore neglect the higher-order terms and consider scalar-tensor theories
with 
\begin{equation}\label{conformal_factor}
\phi = \exp(-\beta \varphi^2)\,
\end{equation}
[corresponding to $\omega(\phi)=-3/2-\kappa/(4\beta \log\phi)$],
with the additional requirement that the scalar field $\varphi$ approach $\varphi_0$  far from the system under consideration (i.e. the role of
the FJBD parameter $\omega_0$ is played by the value $\varphi_0$ of the Einstein-frame scalar field near spatial infinity).
Because essentially of Eq.~\eqref{varphi0}, and since solar-system tests bound $\omega_0>40000$, $\varphi_0$ is constrained
to be very close to zero, while $\tilde{\beta}=\beta/( 4\pi G) \gtrsim-4.5$ because of binary-pulsar 
measurements~\cite{ST_pulsars1,ST_pulsars2,ST_pulsars3}. 

In Ref.~\cite{ourselves} we performed an exact integration (up to numerical errors) of the field equations \eqref{einstein} -- \eqref{scalarT} for
close binary NS systems. In this paper, we seek instead an approximate and further elucidating physical description of such systems, hence it
is natural to resort to PN theory, i.e. to expand the field equations in orders of $v/c$, where $v$ is the binary's velocity. PN theory 
describes extended bodies (i.e. NSs in our case) with a point-particle model, and in GR the masses of these particles are 
constant, as
one would intuitively expect. However, the description of analog systems in scalar-tensor theories is more complicated. 

As can be seen from the action \eqref{Jframe_action}, the scalar field $\phi$ multiplies the Ricci scalar, which in GR is only multiplied by $1/(16 \pi G)$. Therefore, it is
not surprising that in scalar-tensor theories the measured value of the gravitational constant depends on the local value of the scalar field $\phi$, i.e.
the  local gravitational constant $G_{N}$ measured by Cavendish-type experiment is related to the bare $G$ appearing in the action by~\cite{nordtvedt,fierz,jordan,BD}
\begin{equation}
G_{N}=\frac{G}{\phi(\varphi_0)} \frac{4+2\omega_0}{3+2\omega_0}\,.
\end{equation}
This in turns implies that the binding energy of a body, being proportional to the gravitational constant, depends on the value of $\phi$, which in general changes with position.
As a result, when representing 
strongly-gravitating bodies (such as NSs and BHs), for which the binding energy provides a significant portion of the gravitational mass, the masses
of the point-particles ($m_i$, where the index $i$ characterizes the particle) cannot be assumed to be constant in scalar tensor theories. More precisely, the dependence of the
masses on the scalar field is parametrized by the ``sensitivities'', which are defined by~\cite{eardley_dipole}
\begin{equation}
	\label{eqn:sens_def}
	s_i = \frac{\partial \ln m_i(\phi)}{\partial \ln\phi} \,,
\end{equation}
where the derivative is taken while keeping the (Jordan-frame) baryonic mass $m_{\rm bar}$ fixed.
For FJBD, the sensitivities scale roughly as the binding energy per unit mass; thus, they are negligible for stars like
the Sun ($s\sim 10^{-6}$) and white dwarfs ($s\sim 10^{-4}$). However, they are significant for NSs ($s\sim 0.2$) 
and for BHs ($s=1/2$)~\cite{eardley_dipole}.

In the context of the scalar-tensor theories that we consider in this work, it is convenient to 
introduce also the scalar charges $\alpha_i$~\cite{damour_farese_long}, 
which are defined as
\begin{equation}\label{defAlpha}
\alpha_i= - \frac{\partial \ln m^E_{i}(\varphi)}{\partial \tilde{\varphi}}\,,
\end{equation}
where $m^E_{i}=m_i/\sqrt{\phi(\varphi)}$ is the mass in the Einstein frame,
and where the derivative is again taken while keeping the Jordan-frame baryonic mass $m_{\rm bar}$ fixed.
(Note also that our $\alpha_i$ differs from the scalar charge used in Refs.~\cite{damour_farese_long,spontaneous_scalarization,spontaneous_scalarization_bis}
by a minus sign.)
From this definition, one
can show that the scalar charges are related to the sensitivities by~\cite{Damour:1995kt,will_latest}
\begin{equation}
	\label{eqn:sens-scChg}
	\alpha_i = - \frac{2 s_i-1}{\sqrt{3+2\omega_0}}\,.
\end{equation}
As we will see, in the last stages of the evolution of NS binaries or close transient encounters,
DS and IS  produce scalar charges that are $\sim 1$ for NSs, even in the limit
$\varphi_0\to0$ (i.e. $\omega_0\to+\infty$), and that corresponds to diverging sensitivities. 

Modeling therefore a binary system with point particles having masses that depend on the local value of the scalar field,
and performing a PN expansion of the field equations in the ratio between the binary's velocity 
and the speed of light, one finds, after laborious calculations, the PN equations of motion 
for each binary's component~\cite{damour_farese_long,will_zaglauer,will_latest}. In particular, in terms of the binary's separation $\mathbf{x} = \mathbf{x}_1 - \mathbf{x}_2$,
the equations of motions through 2.5 PN order take the schematic form~\cite{will_latest}
\begin{eqnarray}
	\label{eqn:ST-2.5pn}
&&	\frac{d^2 \mathbf{x}}{dt^2} = -\frac{G_{\rm eff} M}{r^2} \mathbf{n} 	\nonumber\\
			&& + \frac{G_{\rm eff} M}{r^2} \left[\left(\frac{\mathcal{A}_{PN}}{c^2}+\frac{\mathcal{A}_{2PN}}{c^4}\right)\mathbf{n}\
			   + \left( \frac{\mathcal{B}_{PN}}{c^2} + \frac{\mathcal{B}_{2PN}}{c^4} \right) \dot{r} \bf{v} \right]	\nonumber\\
			&&	+ \frac{8}{5} \eta \frac{\left(G_{\rm eff} M\right)^2}{r^3} \left[\left( \frac{\mathcal{A}_{1.5PN}}{c^3} +\frac{ \mathcal{A}_{2.5PN}}{c^5}\right) \dot{r} \mathbf{n} \right.	\nonumber\\
			&&	\left.- \left( \frac{\mathcal{B}_{1.5PN}}{c^3} + \frac{\mathcal{B}_{2.5PN}}{c^5}\right) \mathbf{v}\right]
\end{eqnarray}
where 
$M=m_1+m_2$ is the total mass of the system, $\eta = (m_1 m_2)/M^2$ is the symmetric mass ratio,
$r = |\mathbf{x}|$, $\mathbf{n} = \mathbf{x}/r$,
$\mathbf{v}=\mathbf{v}_1 - \mathbf{v}_2$ is the relative velocity of the system,
and $\dot{r} = dr/dt$. The first term on the right-hand side is
the Newtonian acceleration, but the ``effective''
gravitational constant $G_{\rm eff}$ that appears in it is related to
the local gravitational constant $G_{N}$ by
\begin{multline}\label{Geff}
G_{\rm eff}=G_N \left[\frac{3+2 \omega_0}{4+2\omega_0}+\frac{(1-2 s_1)(1-2 s_2)}{4+2\omega_0}\right]\\=G_N 
\left[1+\alpha_1\alpha_2+{\cal O}\left(\frac{1}{\omega_0}\right)\right]\,,
\end{multline}
i.e. the scalar charges tend to enhance the gravitational pull between the two stars. The second group 
of terms are the (conservative) 1PN and 2PN corrections to the Newtonian dynamics, and the third group of terms are 
the dissipative corrections that account of the backreaction of gravitational-wave emission.
Note that dissipative effects appear already at 1.5PN order in the equation of motion, while they only appear
at 2.5PN order in GR. This is because the sensitivities actually source the emission of dipolar
scalar radiation with energy flux
\begin{equation}\label{dipole_flux}
\dot{E}_{\rm dipole} = \frac{G_N}{3 c^3} \left(\frac{G_{\rm eff} m_1 m_2}{r^2}\right)^2 (\alpha_1-\alpha_2)^2+{\cal O}\left(\frac{1}{\omega_0}\right)\,,
\end{equation}
which is 
potentially larger than the usual quadrupolar emission of GR. The explicit expressions for the coefficients $\mathcal{A}_{PN}$, $\mathcal{B}_{PN}$,
$\mathcal{A}_{2PN}$, $\mathcal{B}_{2PN}$, $\mathcal{A}_{1.5PN}$, $\mathcal{B}_{1.5PN}$ , $\mathcal{A}_{2.5PN}$ and $\mathcal{B}_{2.5PN}$
can be found in Ref.~\cite{will_latest}, and depend on the sensitivities of the two stars. It should be noted, however, that
the expressions presented in Ref.~\cite{will_latest} also depend on the first and second derivatives of the sensitivities and of
$\omega$ with respect to the scalar field, evaluated at the asymptotic value $\varphi_0$. Those terms appear because Ref.~\cite{will_latest} expresses the scalar charges
evaluated at the local value of the scalar field as a Taylor expansion around the scalar field $\varphi_0$ at spatial infinity. In fact,
it is possible to write the Lagrangian regulating the motion of binary systems in scalar-tensor theories simply in terms
of the scalar charges $\alpha(\varphi)$ evaluated at the local scalar field~\cite{Damour:1995kt}. Because, as we will explain
in the next section, the formalism that we present in this paper provides directly the scalar charges $\alpha(\varphi)$,
we need to use that information in the equations of motion expressed in terms of $\alpha(\varphi)$ alone, i.e. without performing 
the Taylor expansion of Ref.~\cite{will_latest}. Because, as far as we are aware, 
the equations of motion in terms of $\alpha(\varphi)$ have not been derived explicitly through 2.5 PN order, we reconstruct them
by setting to zero all the terms depending on derivatives of the sensitivities and of $\omega$ in the expressions for the coefficients 
$\mathcal{A}_{PN}$, $\mathcal{B}_{PN}$,
$\mathcal{A}_{2PN}$, $\mathcal{B}_{2PN}$, $\mathcal{A}_{1.5PN}$, $\mathcal{B}_{1.5PN}$ , $\mathcal{A}_{2.5PN}$ and $\mathcal{B}_{2.5PN}$
presented by Ref.~\cite{will_latest}.\footnote{We thank Gilles Esposito-Farese for noting this subtlety. We stress, however, that
the results and conclusions presented in this paper remain qualitatively unchanged if 
the terms depending on derivatives of the sensitivities and of $\omega$ are kept in the expressions of Ref.~\cite{will_latest}.}

Because the motion of a binary system depends on the sensitivities/scalar charges, which
are non-zero for strongly gravitating objects (i.e. ones for which the binding energy 
is not negligible with respect to the gravitational mass) and are in general different for different bodies,
the strong-equivalence principle is violated in scalar-tensor theories. We recall that the equivalence principle
states the universality of free fall for strongly gravitating bodies (in its strong version) 
or for weakly gravitating ones (in its weak version). Clearly, free fall is not universal in scalar tensor theories 
due to the presence  of the sensitivities/scalar charges in the equations of motion (this effect is known
as ``Nordtvedt effect''~\cite{Nordtvedt:1968qr,Roll:1964rd,eardley_dipole}, and takes place generically in the presence of
gravitational degrees of freedom coupled non-minimally to the metric, see e.g. Refs.~\cite{foster1,foster2,yunes}). However, 
the weak version of the equivalence principle is satisfied because the sensitivities
go to zero for weakly gravitating bodies.

In the rest of this paper, we will devise a formalism to calculate the scalar charges/sensitivities for a close binary NS system, simply
by solving a system of algebraic equations, and use them in the PN equations of motion \eqref{eqn:ST-2.5pn}. 
As we will show, our framework allows us to take into account the changes in the scalar charges
during the system's evolution due to the DS and IS, which were discovered with fully relativistic simulations in Ref.~\cite{ourselves}. Therefore, our 
formalism generalizes purely PN approaches such as those of Refs.~\cite{damour_farese_long,ST_pulsars1,ST_pulsars2,ST_pulsars3,will_zaglauer,will_latest}, which assume constant or mildly varying scalar charges/sensitivities and thus
do not account for the effects of DS and IS at small binary separations.

\section{Dynamical Scalarization in Binary Neutron Star Systems}
\label{sec:instantaneous}

As already mentioned, in scalar tensor theories where $\tilde{\beta}$
is sufficiently negative, a non-trivial vacuum for the scalar field develops inside
sufficiently compact isolated NSs, i.e. the scalar field undergoes a ``phase transition''
that is known as ``spontaneous scalarization''~\cite{spontaneous_scalarization,spontaneous_scalarization_bis}.
This phenomenon can be studied in detail by solving the generalized Tolman-Oppenheimer-Volkoff (TOV) equations
governing the structure of isolated NSs in these theories~\cite{spontaneous_scalarization,spontaneous_scalarization_bis}. In particular, 
as proven in Ref.~\cite{damour_farese_long} (Appendix A), the scalar charge of
a NS (formally defined by Eq.~\eqref{defAlpha}) can be extracted from the behavior of the scalar field
near spatial infinity, i.e.
\begin{equation}\label{phi_asymptotic}
\varphi=\varphi_0+\frac{\varphi_1}{r}+  {\cal O} \left(\frac{1}{r^2}\right),
\end{equation}
using the following expression~\cite{damour_farese_long}
\begin{equation}\label{alpha-phi1}
\alpha={\sqrt{4\pi G}}\,\frac{\varphi_1}{\ell_{E}}\,,
\end{equation}
where $\ell_{E}$ is a length scale defined by the asymptotic expansion $g^{E}_{tt}=-1+2 \ell_{E}/r+...$ (i.e. $\ell_{E}$ is proportional to the mass
of the star in the Einstein frame). One can therefore obtain 
the scalar charge $\alpha$ as a function of $\tilde{\beta}$, the asymptotic value $\varphi_0$ of the scalar field,
and the  compactness $C= \ell/R$ of the star, where the lenghtscale $\ell$ is defined 
by the asymptotic expansion $g_{tt}=-1+2 \ell/r+...$ of the Jordan frame metric near spatial infinity. Note that
the gravitational mass $m$
of the star [which is related to $\ell$ through $\ell=G_N [1-s/(\omega_0+2)] m$~\cite{will_zaglauer}] and its radius $R$ are \textit{not} independent, once an equation of state for the NS material has been chosen.
Here, for concreteness, we adopt the same polytropic equation of state 
as in Ref.~\cite{ourselves}, i.e. we choose $K=123 G^3 M_\odot^2/c^6$
and $\Gamma=2$, which yields a maximum mass $m\approx1.8 M_\odot$ both in GR and in the 
scalar-tensor theories we consider, and which provides a reasonable approximation for the equation of state of cold NSs. (Nevertheless
we stress that the procedure
outlined in our model is general, and can be easily extended to any relevant equation of state.)

\begin{figure}
\centering
\includegraphics[height=7.0cm,width=8.5cm,angle=0]{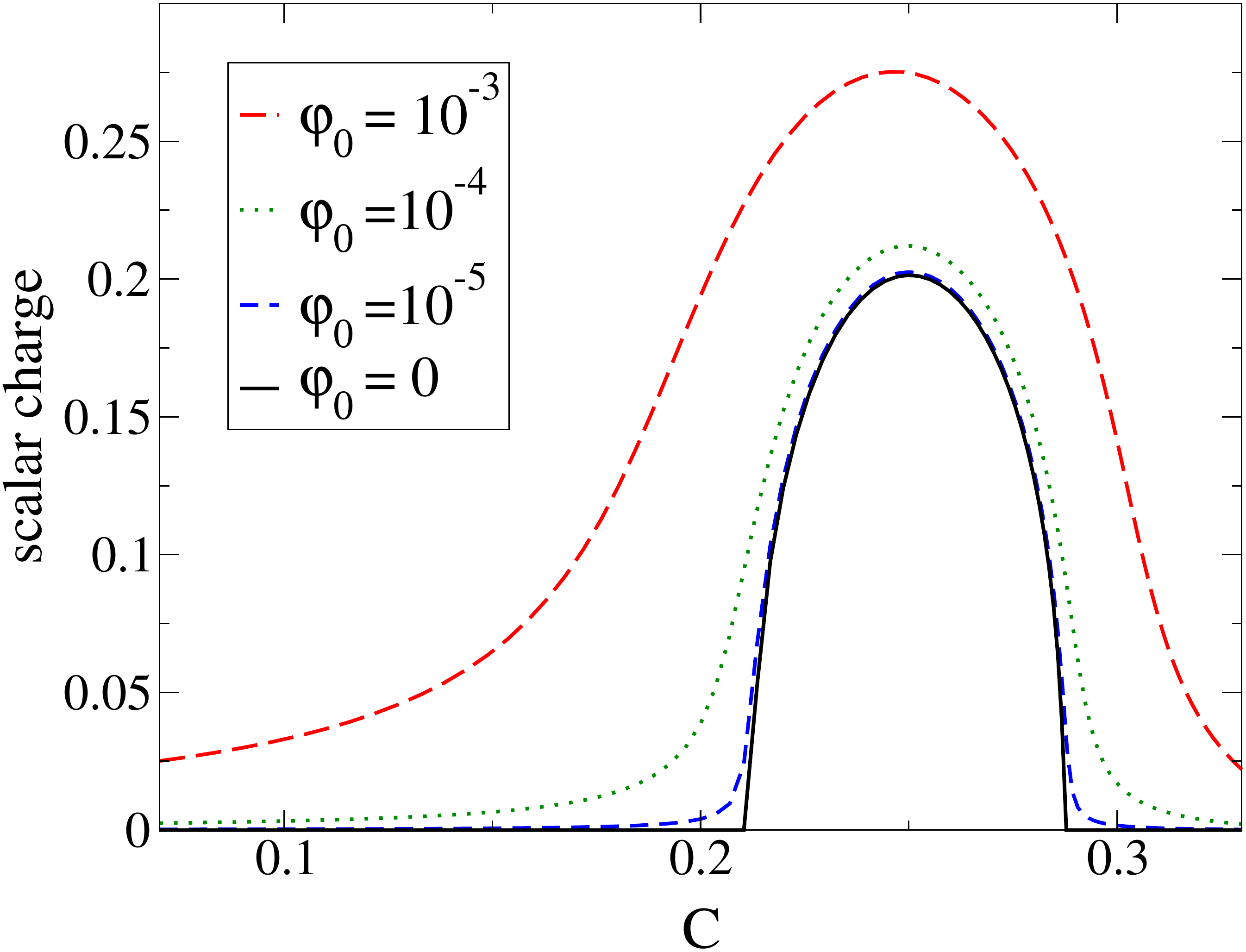}\\
\vskip 0.5cm \hskip 0.0 cm
\includegraphics[height=7.0cm,width=8.5cm,angle=0]{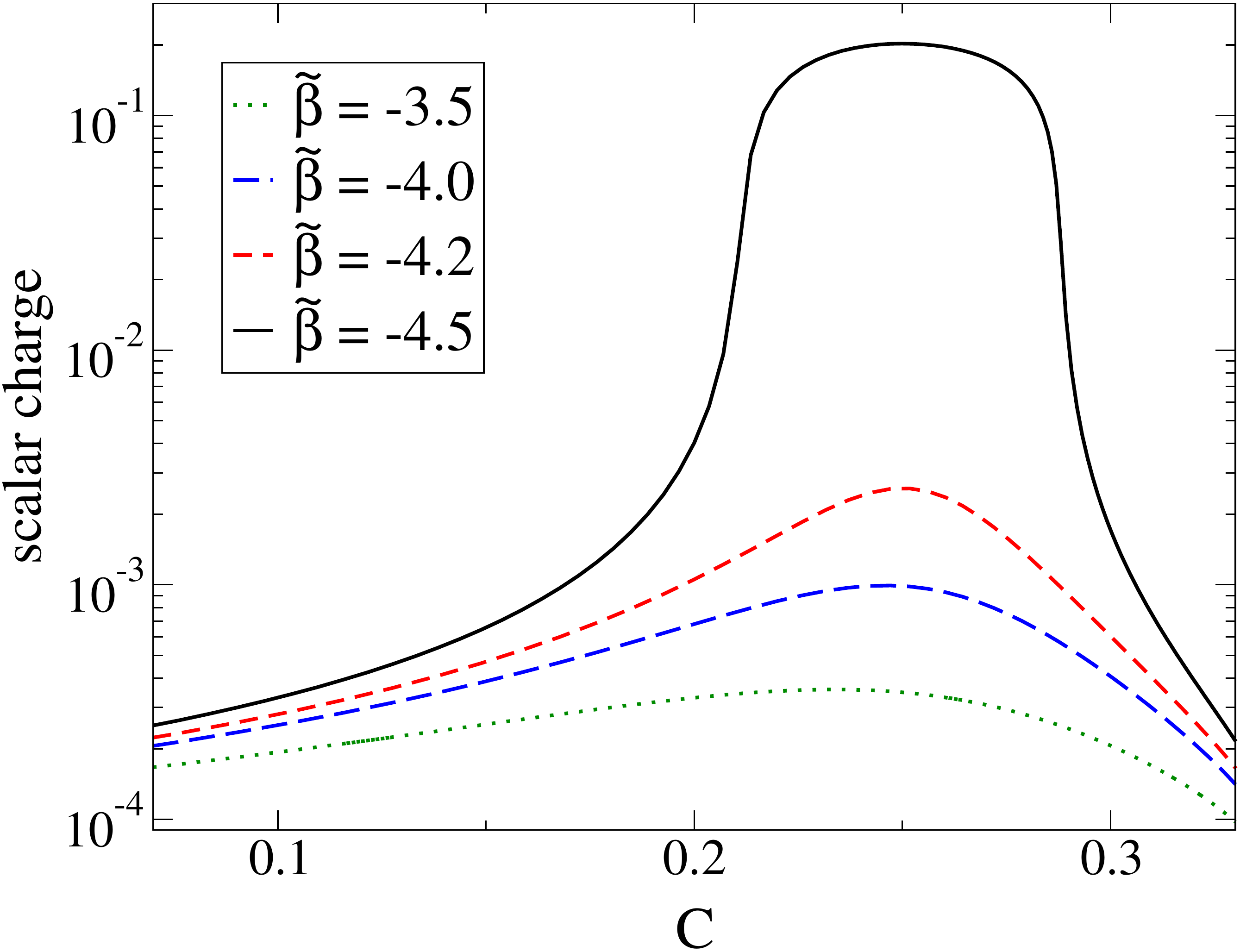}\\
\caption{\footnotesize The scalar charge for an isolated NS,
as a function of compactness,
for different values of $\varphi_{0}$ with fixed $\tilde{\beta}=-4.5$ (top)
and for different values of $\tilde{\beta}$ with fixed $\varphi_{0} G^{1/2}=10^{-5}$
(bottom). As the top panel indicates, spontaneous scalarization occurs at $C \approx 0.21$ for
$\tilde{\beta} \approx -4.5$.
\label{fig:alpha_singlestar}}
\end{figure}

Results for the scalar charge are shown in Fig.~\ref{fig:alpha_singlestar}. In the top panel, as an
illustration, we
consider a theory with $\tilde{\beta}=-4.5$ and various values 
of $\varphi_0$, as a function of the NS compactness. As can be seen, for  $\varphi_0=0$
a sharp discontinuity (corresponding to the spontaneous scalarization mentioned above) 
develops at a critical compactness $C_*\approx0.21$. This sharp transition gets
increasingly blurred as $\varphi_0$ increases, in agreement 
with the results of Ref.~\cite{spontaneous_scalarization,spontaneous_scalarization_bis}.
Note that the maximum $\varphi_0$ allowed by solar-system tests is given by 
Eq.~\eqref{varphi0} with $\omega_0=40000$, and that for $\tilde{\beta}\approx -4.5$ the bound is even tighter~\cite{ST_pulsars1,ST_pulsars2,ST_pulsars3}. However,
as we pointed out in Ref.~\cite{ourselves}, the ``effective'' $\varphi_0$ relevant for stars in a binary system can be much larger then the
solar-system limit.

We also stress that for $\varphi_0=0$, there are actually two families of solutions.
The first corresponds to non-scalarized stars whose structure is the same as in GR
and which have $\alpha=0$. Solutions in this family exist for arbitrary compactnesses $C$, but become
unstable for $C>C_*$, where a second branch of solutions appears, corresponding
to the scalarized stars with $\alpha\neq0$ shown in Fig.~\ref{fig:alpha_singlestar} (in the top panel).
These other solutions are stable (at least until they become too compact and collapse to BHs, which takes place
at compactnesses $C\approx0.25$).
For $\varphi_0\neq0$, instead, the GR-like $\alpha=0$ solutions still exist for 
arbitrary compactnesses $C$, but they are always unstable, while a second 
branch of scalarized solutions, shown in Fig.~\ref{fig:alpha_singlestar} (in the top panel),
exists for arbitrary compactnesses (unlike in the $\varphi_0=0$ case, where they are only present for $C>C_*$). 
These solutions are again stable at least for $C\lesssim 0.25$.
The bottom panel of Fig.~\ref{fig:alpha_singlestar} shows the scalar charge as function of $\tilde{\beta}$ 
and the stellar compactness, for a fixed value of $\varphi_0 G^{1/2}=10^{-5}$. As can be seen,
spontaneous scalarization is important only for $\tilde{\beta}\approx-4.5$, a value almost ruled out by binary-pulsar observations.
[Note that the precise lowest allowed value for $\tilde \beta $ depends
 on the equation of state assumed to describe NSs (see e.g. Ref.~\cite{shibata_buonanno}). In this work, we will
take $\tilde \beta  \geq -4.5$ for concreteness, in order to describe general properties of binary systems within scalar-tensor theories.]

As mentioned in the previous section, in Ref.~\cite{ourselves}
we uncovered, through numerical evidence and analytical arguments, that a ``phase transition'' akin to spontaneous scalarization
takes place in dynamical contexts.
In particular, we showed that in binary NS 
systems (which are brought to sufficiently small separations by the loss of energy and angular momentum through gravitational waves),
a feedback mechanism appears that can induce scalarization in one or
both stars. We refer to this process as dynamical scalarization, and subsequent numerical confirmation of its existence 
has been recently brought forward by Ref.~\cite{shibata_buonanno}.

As we discuss next, DS is similar to
spontaneous scalarization, but is regulated by the ``effective'' value of the background scalar 
field felt by each star in the binary, as well 
by the ``effective'' compactness of the system.
The former case (i.e. DS regulated by the ``effective'' value of the scalar field) is most relevant in binaries 
consisting of a scalarized star 
and a non-scalarized one. In this case, DS amounts to a non-linear version of IS~\cite{ourselves,Ruiz:2012jt}, where the 
non-scalarized star is immersed in an increasingly large scalar-field background produced by the scalarized star. Therefore,
the unscalarized star 
acquires a scalar charge, in agreement with the top panel of Fig.~\ref{fig:alpha_singlestar}, which shows that $\alpha$ grows with 
$\varphi_0$, even for stars with negligible scalar charge for  $\varphi_0\to 0$. An important observation is that this process is \textit{non-perturbative} 
because the newly-scalarized star will in turn induce a growth in the scalar charge of its companion. In other words, 
a dynamical interplay arises because either 
star (say star 1) is not only sensitive to the asymptotic value of the scalar field far away from the binary, but also to the contribution
due to its companion (say star 2), hence
\begin{equation}\label{background_field1}
  \varphi_{B}^{(1)} = \varphi_{0} + \frac{\varphi_1^{(2)} ( \varphi_{B}^{(2)})}{r} + {\cal O} \left(\frac{1}{r^2}\right),
\end{equation}
where $ \varphi_{B}^{(i)}$ ($i=1,2$) is the ``background'' value of the
scalar field in which star $i$ is immersed, and we recall that the
coefficient $\varphi_1^{(j)} ( \varphi_{B}^{(j)})$ [defined by the asymptotic expansion \eqref{phi_asymptotic}] 
is proportional to the scalar charge [cf. Eq.~\eqref{alpha-phi1}], and thus a function of the scalar-field background
in which the star is immersed (cf. Fig.~\ref{fig:alpha_singlestar}, top panel). Clearly, a similar equation will describe the interaction of star 2 with the background
scalar field produced by star 1
\begin{equation}\label{background_field2}
  \varphi_{B}^{(2)} = \varphi_{0} + \frac{\varphi_1^{(1)} ( \varphi_{B}^{(1)})}{r} + {\cal O} \left(\frac{1}{r^2}\right) \, .
\end{equation}
The system \eqref{background_field1}--\eqref{background_field2} then describes
 the feedback mechanism alluded above.\footnote{We stress that the uncontrolled remainders ${\cal O} ({1}/{r^2})$ appearing in Eqs.~\eqref{background_field1} and \eqref{background_field2}
can be safely neglected, as we have verified explicitly that extracting them 
from our TOV isolated-star solutions and including them in Eqs.~\eqref{background_field1} and \eqref{background_field2} has a negligible effect on our results.
One can easily make sense of why that must be the case. In fact, the ${\cal O} ({1}/{r^2})$ terms 
introduce a correction of $\sim 3-4\%$ on the value of $\varphi$ at the separations of $50-60$ km where the plunge takes place.
At smaller separation the correction will be larger, but \textit{(i)} the effect of the scalar charges is negligible during the plunge, since that
happens on the dynamical timescale, and \textit{(ii)} even at the closest separation $r\sim 2 R$, where the two stars merge, the effect of
those terms on the value of $\varphi$ is only $\sim 7-8\%$.}
 In practice, we model the functions 
$\varphi_1^{(i)} ( \varphi_{B}^{(i)})$ ($i=1,2$) with a fit to data for $\varphi_1$ coming from solutions to the generalized-TOV equations~\cite{spontaneous_scalarization}
describing an isolated NS, for various values of the scalar field $\varphi_0$ at spatial infinity. Equations \eqref{background_field1}--\eqref{background_field2}
then become a system of (non-linear) algebraic equations which one can easily solve numerically. For concreteness, one method of solving this system 
is to look iteratively for a fixed point. 
At the first iteration, this method
therefore yields the same results as the IS described e.g. by Ref.~\cite{Ruiz:2012jt}, but at the following ones the feedback
mechanism described above becomes important. Therefore, IS qualitatively accounts for the scalarization of non-scalarized stars
that get close to scalarized ones, but fails to describe for the non-perturbative feedback exerted by the newly-scalarized star on its companion.

In principle, the iteration may not converge to a fixed point, or even worse the system \eqref{background_field1}--\eqref{background_field2}
may not have any real solutions. It is straightforward, however, to show that the fixed point method indeed converges to a solution.
Let us consider a scalar tensor theory with given $\tilde{\beta}$ and $\varphi_0$, and two stars with (Jordan-frame) baryonic masses $m^{\rm bar}_i$ at separation $r$.
(Note that the Jordan-frame baryonic masses are conserved, see e.g. Ref \cite{ourselves}). Our algorithm then  proceeds as follows
\begin{enumerate}
\item At the initial iteration,
we set $\varphi_B^{(i)} = \varphi_{0}$ if the stars are widely separated, or to a better guess. For example, if the $\varphi_B^{(i)}$
have been already calculated for a nearby separation (e.g. at a previous close instance of the dynamics), we may assume those values as the starting point of our iteration. Similarly, starting from
the second iteration, we set $\varphi_B^{(i)}$ to the values produced by the previous iteration.

\item For each star, we use a code solving the generalized-TOV equations~\cite{spontaneous_scalarization}
to find the parameter $\varphi_1^{(i)}$ for an isolated NS with
given baryonic mass  $m^{\rm bar}_i$ in a scalar tensor theory where the asymptotic value of the scalar field is set to the value $\varphi_B^{(i)}$ of the background scalar field. (Note that 
$\tilde{\beta}$ is fixed.) In practice, to speed up this step, we produce data for  $\varphi_1$ as a function of baryonic mass and asymptotic scalar field value,
and fit them in the neighborhood of the target values $m^{\rm bar}_i$ and $\varphi_B^{(i)}$.

\item Next, we update the background scalar field value via Eqs.~\eqref{background_field1} -- \eqref{background_field2}.

\item Steps 1-3 are iterated until the solution is found, e.g. until the relative
difference between $\varphi_B^{(i)}$ at consecutive steps drops below
a given tolerance. 

\item The scalar charges can be obtained from the final values of  $\varphi_1^{(i)}$ 
with Eq.~\eqref{alpha-phi1}, using the TOV code (or a fit to its results) to compute $\ell_E$.

\end{enumerate}

Convergence of this method requires that the ratio
between the variations of $\varphi_B^{(i)}$ at iteration $n+1$ and $n$
be $\leq 1$.  It is easy to show (see the Appendix for details) that 
this  condition is satisfied if
\begin{equation}\label{stability_solution}
   \frac{1}{r} \sqrt{ \frac{\partial {\varphi}_1^{(1)}}{\partial \varphi^{(1)}_B}
                      \frac{\partial {\varphi}_1^{(2)}}{\partial \varphi^{(2)}_B}
                     } \le 1  .
\end{equation}
We monitor this condition as the iterations proceed and confirm
it is typically well below the bound.~\footnote{Interestingly, we find that
the only separation $r$ at which this bound is approached (but never violated)
is the one marking the onset of DS, i.e. the one
corresponding to the critical ``effective'' compactness  $\bar{C}_*$ defined later in this section.}
Note that alternatively one can solve the system \eqref{background_field1}--\eqref{background_field2}
(again, with the functions $\varphi_1^{(i)}$ obtained as fits to TOV data) directly, with a
two-dimensional Newton-Raphson method. 
This method leads to the same solution as the fixed-point method.

A situation where the feedback mechanism described by Eqs.~\eqref{background_field1}--\eqref{background_field2} 
is particularly important is that involving two stars that have exactly zero scalar charge in isolation. This would
be the case, for instance, for a theory with $\varphi_0=0$ and for stars with $C<C_*$ (cf. Fig.~\ref{fig:alpha_singlestar}). In the ``perturbative'' picture of IS,
no scalar charges should develop in such a situation. That would indeed correspond to the trivial solution $\varphi_{B}^{(i)}=0$ ($i=1,2$) for the system 
\eqref{background_field1}--\eqref{background_field2}. However, in Ref.~\cite{ourselves} we showed with fully relativistic numerical simulations
that even in such a situation the NSs scalarize, when the binary's separation shrinks to a sufficiently small value. What happens physically is
similar to the spontaneous scalarization of isolated stars, i.e. because of the
feedback mentioned above, the system \eqref{background_field1}--\eqref{background_field2} develops a non-trivial solution (i.e. one with $\varphi_{B}^{(i)}\neq0$) when the effective ``compactness''
of the binary, defined as $\bar{C}=G_N E_{\rm tot}/r$ (where $E_{\rm tot}$ is the total energy of the system -- including the two masses -- and $r$ is the separation) reaches a critical threshold. 

\begin{figure}
\centering
\includegraphics[height=7.0cm,width=8.5cm,angle=0]{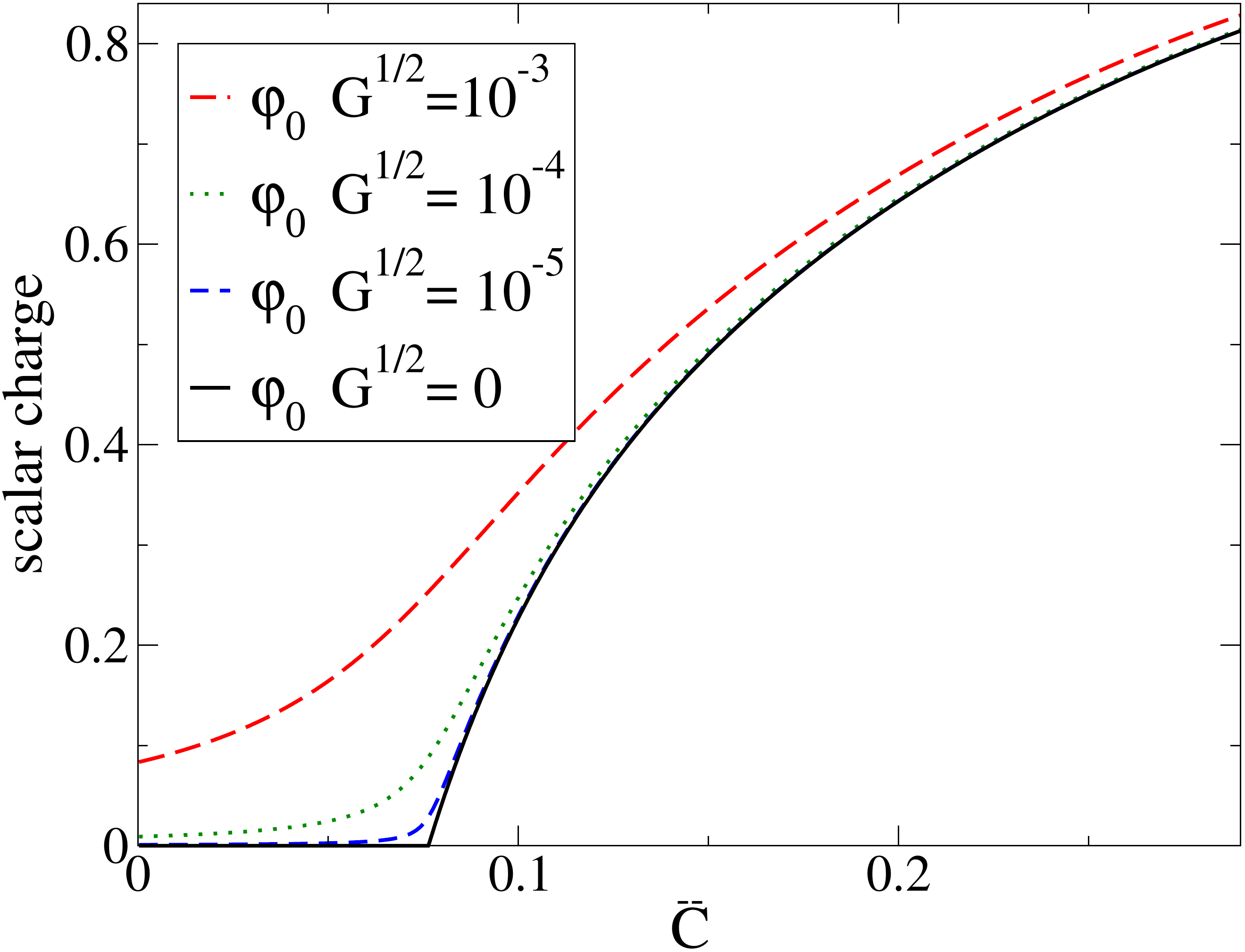}\\
\vskip 0.5cm \hskip 0.0 cm
\includegraphics[height=7.0cm,width=8.5cm,angle=0]{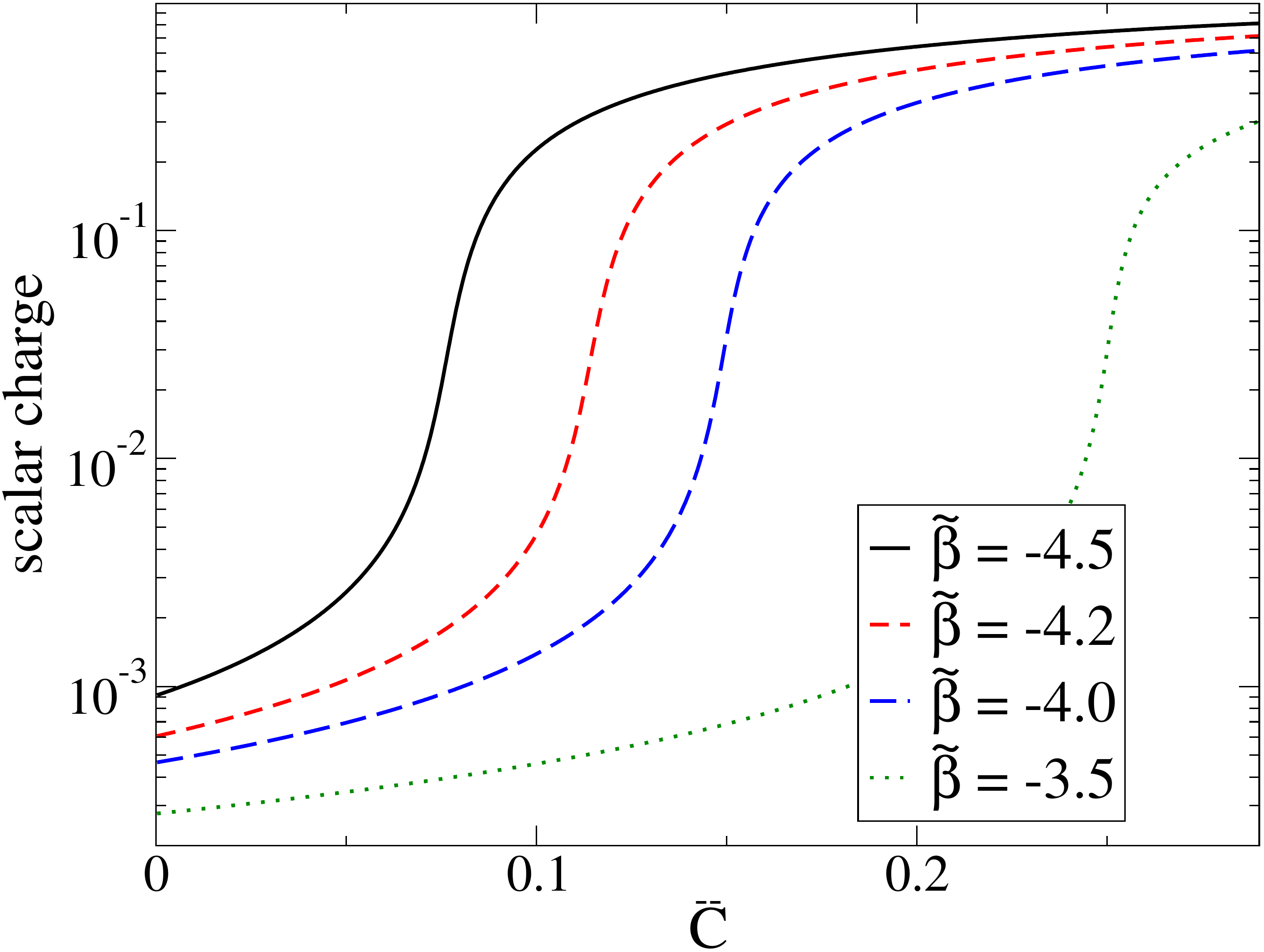}
\caption{\footnotesize Scalar charge as a function of the binary's compactness
$\bar{C}=G_N E_{1PN}/r$, for an equal-mass system with $m_1=m_2=1.51 M_{\odot}$,
for different values of $\varphi_{0}$ with fixed $\tilde{\beta}=-4.5$ (top)
and for different values of $\tilde{\beta}$ with fixed $\varphi_{0} G^{1/2} =10^{-5}$
(bottom). Note that DS occurs for any value
of $|\tilde{\beta}|$, although the critical compactness $\bar{C}_*$
is shifted to higher values as  $|\tilde{\beta}|$ decreases.
\label{fig:alpha_binary}}
\end{figure}

Results for scalar charge obtained by solving Eqs.~\eqref{background_field1}--\eqref{background_field2}
in one such case, namely for an equal-mass binary $m_1=m_2=1.51 M_{\odot}$ made of stars that do not scalarize
spontaneously in isolation, are shown in 
Fig.~\ref{fig:alpha_binary}, for scalar tensor theories with
 various values of $\tilde{\beta}$ and $\varphi_0$. The scalar charge is plotted as a function
of the effective compactness of the binary, calculated as $\bar{C}=G_N E_{1PN}/r$ (where 
$E_{1PN}$ is the total binary energy at 1PN order~\cite{will_latest}).

The top panel, in particular, shows the case $\tilde{\beta}=-4.5$ 
for various values of $\varphi_0$. As can be seen,
for $\varphi_0=0$,  the scalar charges are exactly zero
at large separations, because the stars have insufficient ``individual'' compactness $C=\ell/R$
to undergo spontaneous scalarization in isolation. However, when the 
effective binary compactness $\bar{C}=G_N E_{1PN}/r$ reaches a critical value $\bar{C}_*\approx 0.75$,
the system does display a phase transition to non-zero charges, i.e. a DS.
This behavior clearly mirrors closely that in Fig.~\ref{fig:alpha_singlestar}
for a star in isolation, and as in that case the DS transition
gets increasingly blurred as $\varphi_0$ increases. Also, again
in agreement with the isolated star case, the system 
\eqref{background_field1}--\eqref{background_field2} still
allows the trivial solution $\alpha_1=\alpha_2=0$, but that solution
is only stable for $\varphi_0=0$ and $\bar{C}<\bar{C}_*$.

The behavior of the scalar charge as a function of $\tilde{\beta}$ is illustrated 
in the bottom panel of
Fig.~\ref{fig:alpha_binary}. As can be observed,
the maximum value of the scalar charge is not very sensitive to  $\tilde{\beta}$, unlike
in the spontaneous scalarization of isolated stars, where $\alpha$ decreases as $|\tilde{\beta}|$ decreases (cf. bottom panel
of Fig.~\ref{fig:alpha_singlestar}). However, the critical
compactness $\bar{C}_*$ at which DS switches on increases as  $|\tilde{\beta}|$ decreases.
Thus, smaller separations are required for the scalar charges to grow. Of course,
when the stars touch each other, DS effects will be subleading relative
to strong material interactions driven by the merger process.

We stress that Fig.~\ref{fig:alpha_binary} is produced
by solving Eqs.~\eqref{background_field1} and \eqref{background_field2}.
While this system is a purely algebraic one, and thus straightforward 
to solve numerically, it neglects the time delay needed for one star to ``feel''
 the change in the scalar field background produced by the other one as it moves, i.e.   
 Eqs.~\eqref{background_field1}--\eqref{background_field2} assume an instantaneous
 feedback between the two stars. We will show in the next section that those equations
 can be coupled to the PN equations of motion to account for the finite propagation speed of the interaction,
but the deviations from the ``instantaneous'' results obtained by solving the simple algebraic system \eqref{background_field1}--\eqref{background_field2} alone
 scale as $\dot{r}/c$. Consequently they are 
negligible during the quasi-adiabatic inspiral of the NS binary, when the separation varies slowly
as function of time.

Because DS provides an efficient and robust mechanism for scalarization 
 -- i.e. it forces generic NS binaries to scalarize at some stage of their
 orbital evolution, even if the individual stars are not compact enough
 to scalarize spontaneously in isolation, and amplifies the effect of IS in systems in which at least one component
 is scalarized in isolation --, it is expected to produce significant deviations from GR
 in the orbital evolution. These deviations are driven by the ensuing enhancement of the gravitational pull
 between the stars [cf. Eq.~\eqref{Geff}] and the emission
 of dipolar scalar waves [cf. Eq.~\eqref{dipole_flux}].
 As was shown in Ref.~\cite{ourselves} (and later confirmed by Ref.~\cite{shibata_buonanno}), these effects
 trigger earlier plunges in NS binaries relative to GR, which could potentially be observable with
 ground-based gravitational-wave detectors.

\begin{figure}
\centering
\includegraphics[height=7.0cm,width=8.5cm,angle=0]{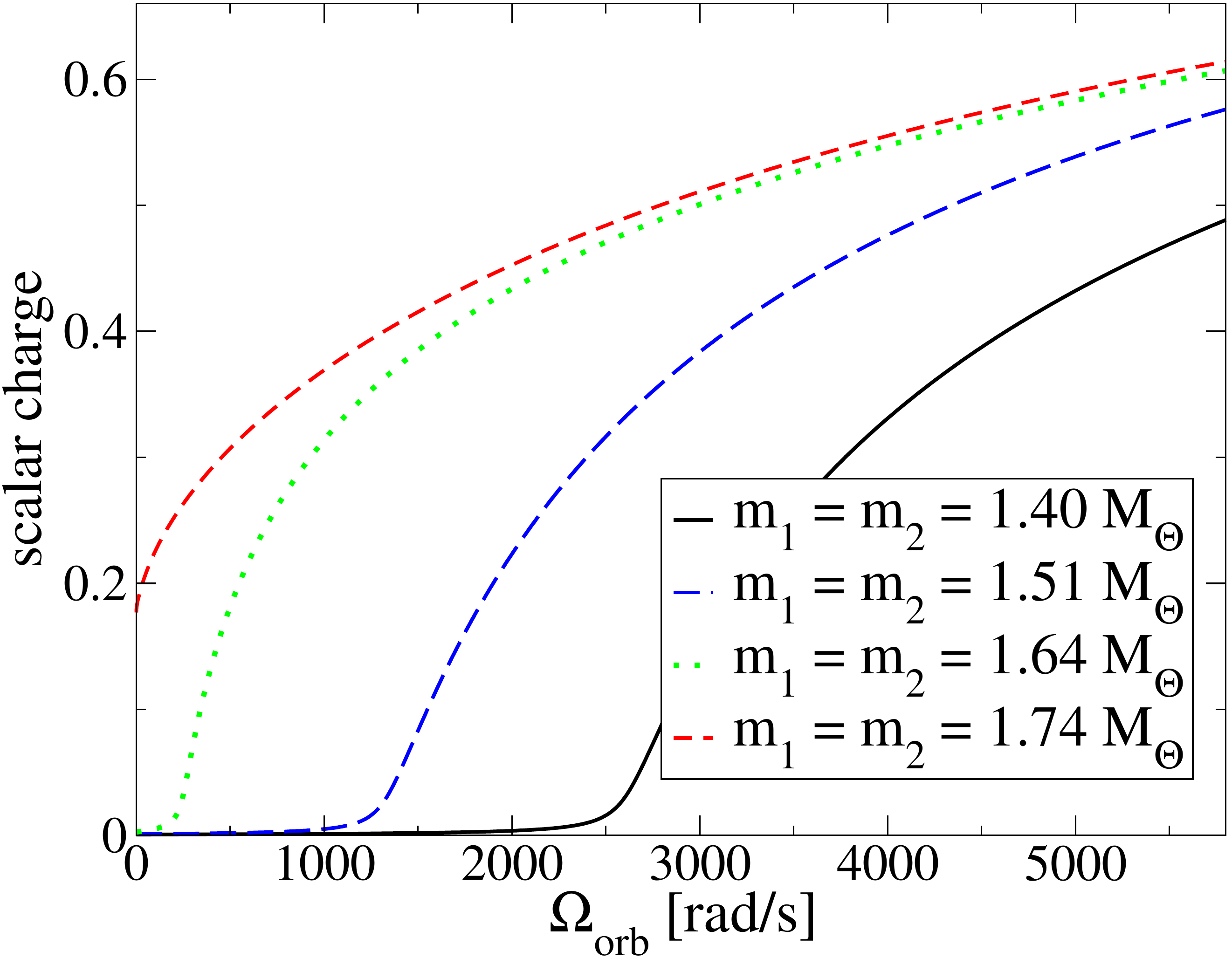}\\
\vskip 0.5cm \hskip 0.0 cm
\includegraphics[height=7.0cm,width=8.5cm,angle=0]{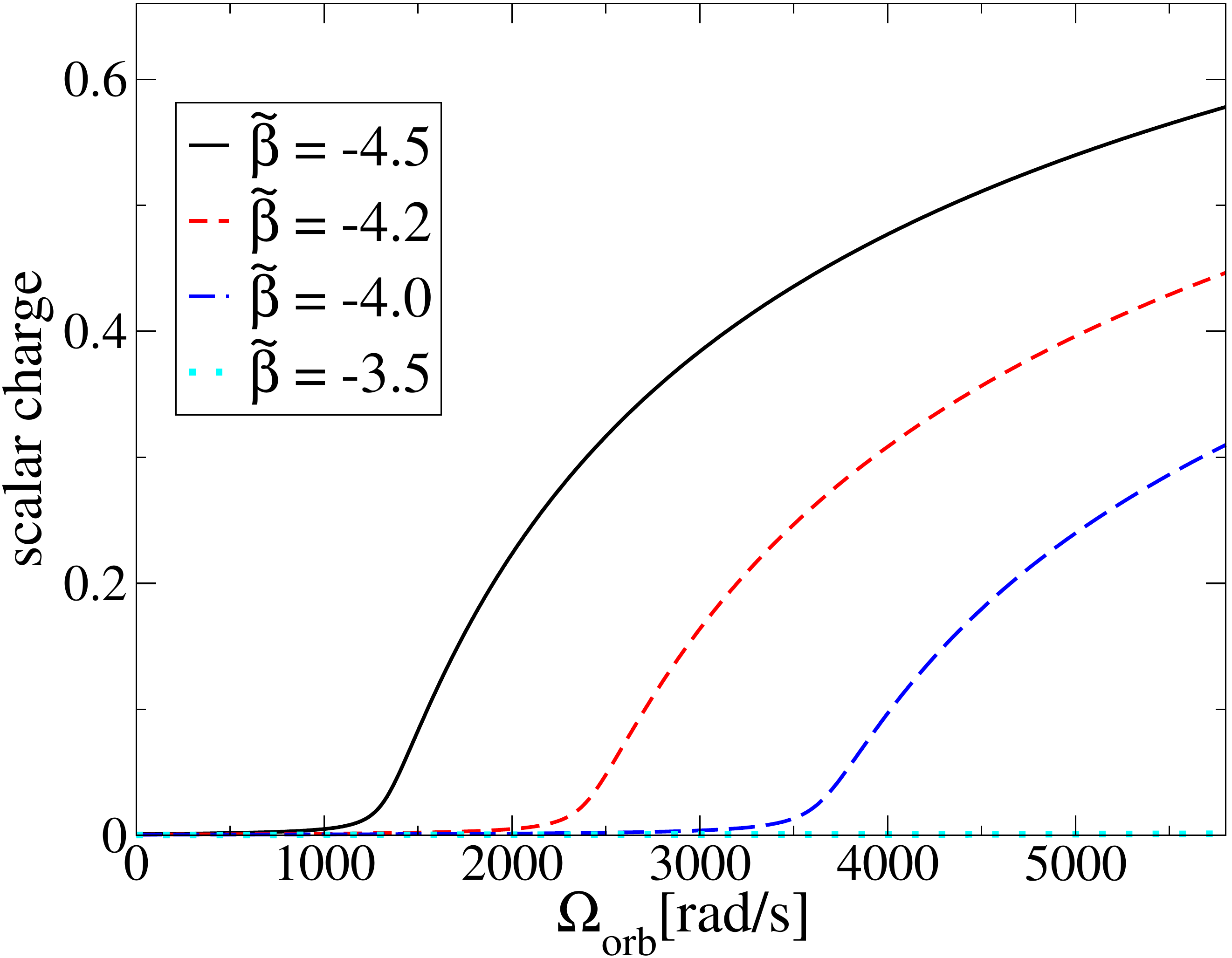}
\caption{\footnotesize Scalar charge as a function of the
orbital frequency of the binary (computed at 2PN order in the quasi-circular approximation) for different equal-mass binaries
with $\varphi_{0} G^{1/2}=10^{-5}$ and $\tilde{\beta} = -4.5$ (top panel) and for an equal-mass
binary $m_1=m_2=1.51 M_{\odot}$ for different values of $\tilde{\beta}$ (bottom).
\label{fig:alpha_omega}}
\end{figure}

Fig.~\ref{fig:alpha_omega} illustrates this effect by plotting the scalar charge
resulting from solution of the system \eqref{background_field1} -- \eqref{background_field2},
for equal-mass NS binaries as a function of the quasi-circular orbital frequency, computed at 2PN order by imposing
$\dot{r}=\ddot{r}=0$ in Eq.~(\ref{eqn:ST-2.5pn}).
The top panel considers several masses for the binary's components
and assumes $\tilde{\beta}=-4.5$ and $\varphi_0 G^{1/2}=10^{-5}$, while the lower panel
considers $m_1=m_2=1.51 M_\odot$, $\varphi_0 G^{1/2} =10^{-5}$ 
and various values of $\tilde{\beta}$. As expected from the previous figures, the onset of DS 
(which, as we will show in the next section and as expected from the arguments above, roughly marks the beginning of the plunge)
moves to larger frequencies as $|\tilde{\beta}|$ decreases or as the NS masses decrease.
(We recall that in GR as well as in scalar tensor theories, lower NS masses correspond to lower compactnesses
$C=\ell/R$.)

The  critical orbital frequency $\Omega_{*}$ marking the onset of DS can be easily fitted as a function
of $m_1$, $m_2$ and ${\tilde \beta}$. (The dependence on $\varphi_0$ is very weak for values allowed by 
solar-system tests.) 
 For instance, for ${\tilde \beta} \in [-4.5,-3.75]$ and
$m_i/M_\odot \in [1.40,1.74]$ the simple expression 
\begin{equation}\label{omega_critical}
   \Omega_{*} [rad/s] = A \left(\frac{m_1}{M_\odot} - m_{*}\right) \left(\frac{m_2}{M_\odot} - m_{*}\right) 
\end{equation}
where
\begin{eqnarray}
  A &=& 690656 +  336637 {\tilde \beta} + 42621 {\tilde \beta}^2 
\nonumber \\
  m_{*} &=& 4.0512 + 0.5123 {\tilde \beta}
\end{eqnarray}
captures (within a maximum error of $15\%$)  the critical frequency for the
polytropic equation of state with $\Gamma=2$ and $K=123 G^3 M_\odot^2/c^6$ that we use in this paper.
Note that the regime ${\tilde \beta} < -4.5$
is ruled out by binary-pulsar observations~\cite{ST_pulsars1,ST_pulsars2,ST_pulsars3}, while  for
${\tilde \beta} > -3.5$ the critical frequency results too high to be reached before the NSs merge.

\section{Orbital Dynamics}

As mentioned above, the solution of the algebraic system \eqref{background_field1} -- \eqref{background_field2}
provides a good description of the scalar charge as a function of orbital frequency during
the adiabatic inspiral phase of the NS binary, but this approximation \textit{(i)} breaks down when the 
binary plunges; \textit{(ii)} does not allow one to calculate self-consistently the impact of the scalar
charges on the orbital evolution. In order to address both limitations, in this section we describe how to complement the algebraic system \eqref{background_field1} -- \eqref{background_field2}
with a set of ordinary differential equations describing a binary's motion at 2.5 PN order.

The equations of motion to 2.5PN order for scalar tensor theories,
given schematically in Eq.~\eqref{eqn:ST-2.5pn},  have been derived recently in Ref.~\cite{will_latest}. 
These equations were derived adopting non-spinning stars and
assumed sensitivities (and thus scalar charges) that evolve only mildly
during the system's evolution\footnote{As mentioned previously, Ref.~\cite{will_latest} allows the sensitivities/scalar charges
to depend on the local value of the scalar field, but expands this dependence in a Taylor series around the asymptotic scalar field $\varphi_0$. This makes it
impossible to account for strong changes of the sensitivities/scalar charges such as those produced by IS and DS.}. The first assumption
is a natural one for NSs, which typically have small spin (see e.g.~\cite{O'Shaughnessy:2009dk}), 
and can be relaxed by including spin-orbit and spin-spin interactions
in the PN equations. The second assumption however, fails to account for the effects 
of DS which, as described,  produces large scalar charges that vary rapidly with frequency (and therefore time)
along the evolution.
Our model accounts for this effect by dynamically adjusting the scalar charges along the evolution, 
according to the feedback
mechanism described in the previous section. However, instead of solving Eqs.~\eqref{background_field1}--\eqref{background_field2}, 
we modify them to account for the retardation effects due to the motion of the binary, i.e. at each timestep we solve the system
\begin{gather}\label{background_field1bis}
  \varphi_{B}^{(1)} = \varphi_{0} + \frac{\varphi_1^{(2)} ( \varphi_{B}^{(2)}  ) }{r(1-\dot{r})} + {\cal O} \left(\frac{1}{r^2}\right),\\
\label{background_field2bis}
  \varphi_{B}^{(2)} = \varphi_{0} + \frac{\varphi_1^{(1)} ( \varphi_{B}^{(1)} )}{r(1-\dot{r})} + {\cal O} \left(\frac{1}{r^2}\right).
\end{gather}
Physically, this system of equations means that at each time $t$, the background scalar field ``felt''
by one star is given by the scalar field exerted by the other star at an \textit{earlier} time $t-r$, when the separation
of the binary was $r(t-r)\approx r(t) (1-\dot{r})$. 
Here, as in the previous section, we model the functions $\varphi_1^{(i)} ( \varphi_{B}^{(i)})$ ($i=1,2$) with  fits to 
data for $\varphi_1$ coming from solutions to the generalized-TOV equations~\cite{spontaneous_scalarization}
describing an isolated NS, for various values of the asymptotic scalar field $\varphi_0$
and fixed Jordan-frame baryonic masses. (Note that the Jordan-frame baryonic masses are conserved during the evolution, 
see e.g. Ref. \cite{ourselves}).
The derivative of the binary's separation, $\dot{r}$,
is instead evaluated at each step with the PN equations of motion \eqref{eqn:ST-2.5pn}. Clearly, in the inspiral $\dot{r}\ll 1$, so
the system \eqref{background_field1bis} -- \eqref{background_field2bis} reduces to Eqs.~\eqref{background_field1}--\eqref{background_field2}.

At a formal level, our model can be thought of as supplementing the PN equations
of motion \eqref{eqn:ST-2.5pn} with two extra equations
\begin{equation}\label{eoscharges}
\frac{d\alpha^i}{dt} =  \frac{d\alpha^{(i)}}{d \varphi_B^{(i)}} \frac{d\varphi_B^{(i)}}{dr} \dot{r}
\end{equation}
describing the evolution of each star's scalar charge.
In these equations, the last term is determined by the stars' trajectories (i.e. by the PN equations of
motion), while $d\alpha/d\varphi_B$ and  $d\varphi_B/dr$ are determined respectively
using the solutions to the generalized TOV equations for isolated stars, and by  
eqs. \eqref{background_field1bis} -- \eqref{background_field2bis}.

\section{Results}
\begin{figure}
\centering
\includegraphics[height=7.0cm,width=8.5cm,angle=0]{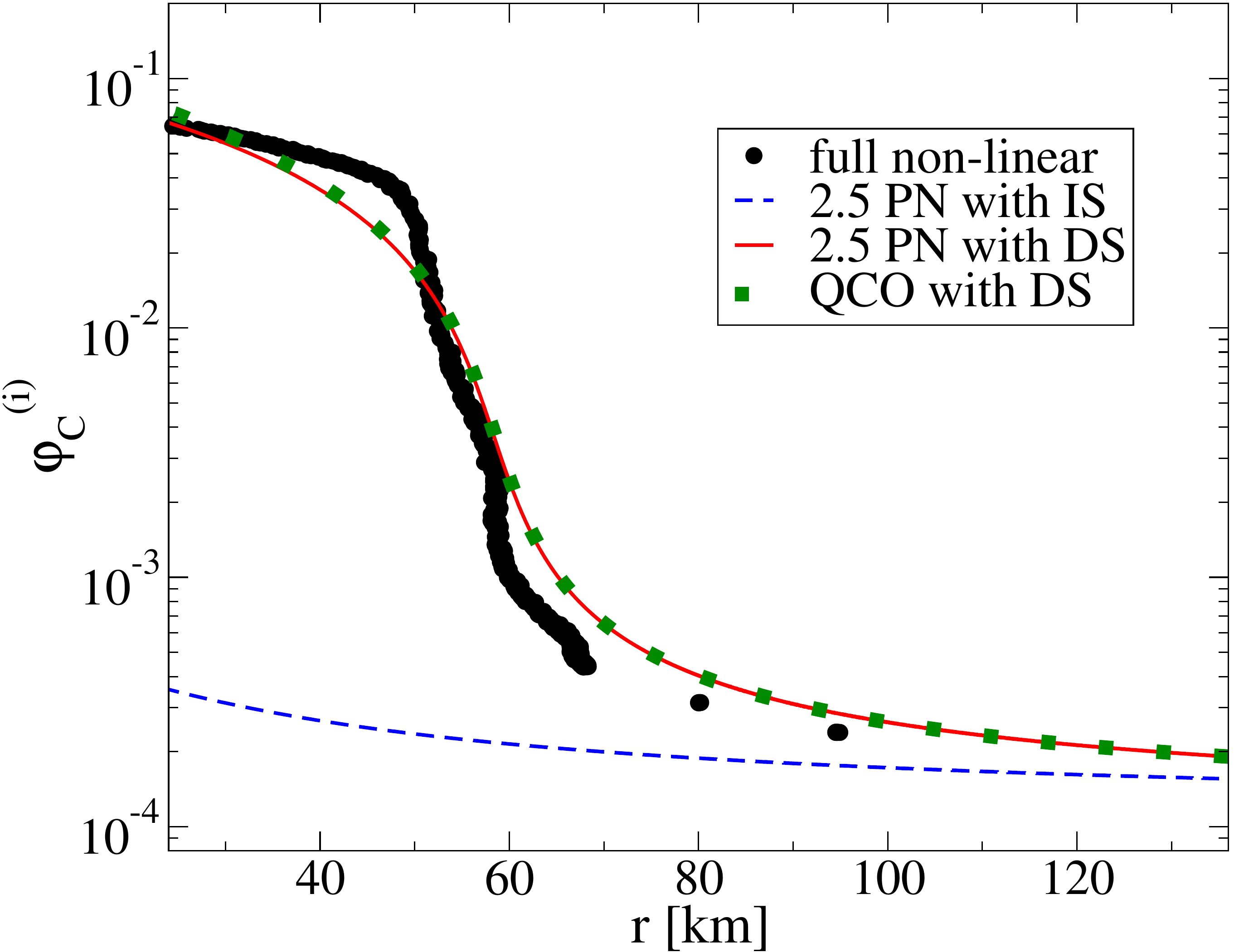}\\
\caption{\footnotesize Central value of the scalar field $\varphi_C^{(1)}=\varphi_C^{(2)}$ 
as a function of the binary separation for the equal mass case
$m_1=m_2=1.51 M_{\odot}$ with ${\tilde \beta} = -4.5$ and $\varphi_{0} G^{1/2} =10^{-5}$.
The different curves correspond to results obtained with: a non-linear simulation (dotted line); 
a 2.5PN evolution accounting for DS (solid line); a 2.5PN evolution accounting only for IS (dashed line).
For reference, we also include results (squares) obtained in the instantaneous quasi-circular orbit (QCO) approximation to DS,
described in Sec.~\ref{sec:instantaneous}.
\label{fig:phic_r_mass15_15}}
\end{figure}

In this section, we summarize the results obtained by evolving the 2.5PN equations of motion
for scalar-tensor gravity with dynamical scalar charges.
For this purpose we have implemented the method described in the previous section, and
integrated the evolution equations by a fourth-order Runge-Kutta solver. This code
has been validated through exhaustive self-consistency tests as well by direct comparison
to known results in the GR limit. Additionally, we can compare with 
our own recent, fully non-linear simulations for binary NS systems in scalar-tensor theories~\cite{ourselves}. 
For these tests, it is cleanest to compare 
the value of the scalar field $\varphi_{C}^{(i)}$ at the center of each star. This value is directly related to the scalar charge.  
Fig.~\ref{fig:phic_r_mass15_15} displays
$\varphi_{C}^{(1)}=\varphi_{C}^{(2)}$ for an equal mass binary with
$m_1=m_2=1.51 M_{\odot}$, in a theory with ${\tilde \beta} = -4.5$ and $\varphi_{0} G^{1/2}=10^{-5}$.
The black dots correspond to the values calculated with a full non-linear evolution from an initial
separation of $70$ km~\cite{ourselves}. Additionally, to extend the reach of this test, we have also 
evolved two larger separations ($80$ km and $100$ km) and extracted the
central value of the scalar field after one orbit (i.e. $\varphi_{C}^{(1)}=\varphi_{C}^{(2)}$ is measured after the initial data have relaxed and 
a quasi-stationary solution is reached). 
We also include in this figure the results obtained with the 2.5PN equations of motions allowing only 
for IS (by not accounting for the DS feedback mechanism described previously), as well as DS. 
As can be seen, the results obtained with DS are
in good agreement with the full non-linear solution, while the ones including only IS
clearly underestimate the growth of the scalar field.
It is also clear that our approach provides a good approximation up to
close separations.

\begin{figure}
\centering
\includegraphics[height=7.0cm,width=8.5cm,angle=0]{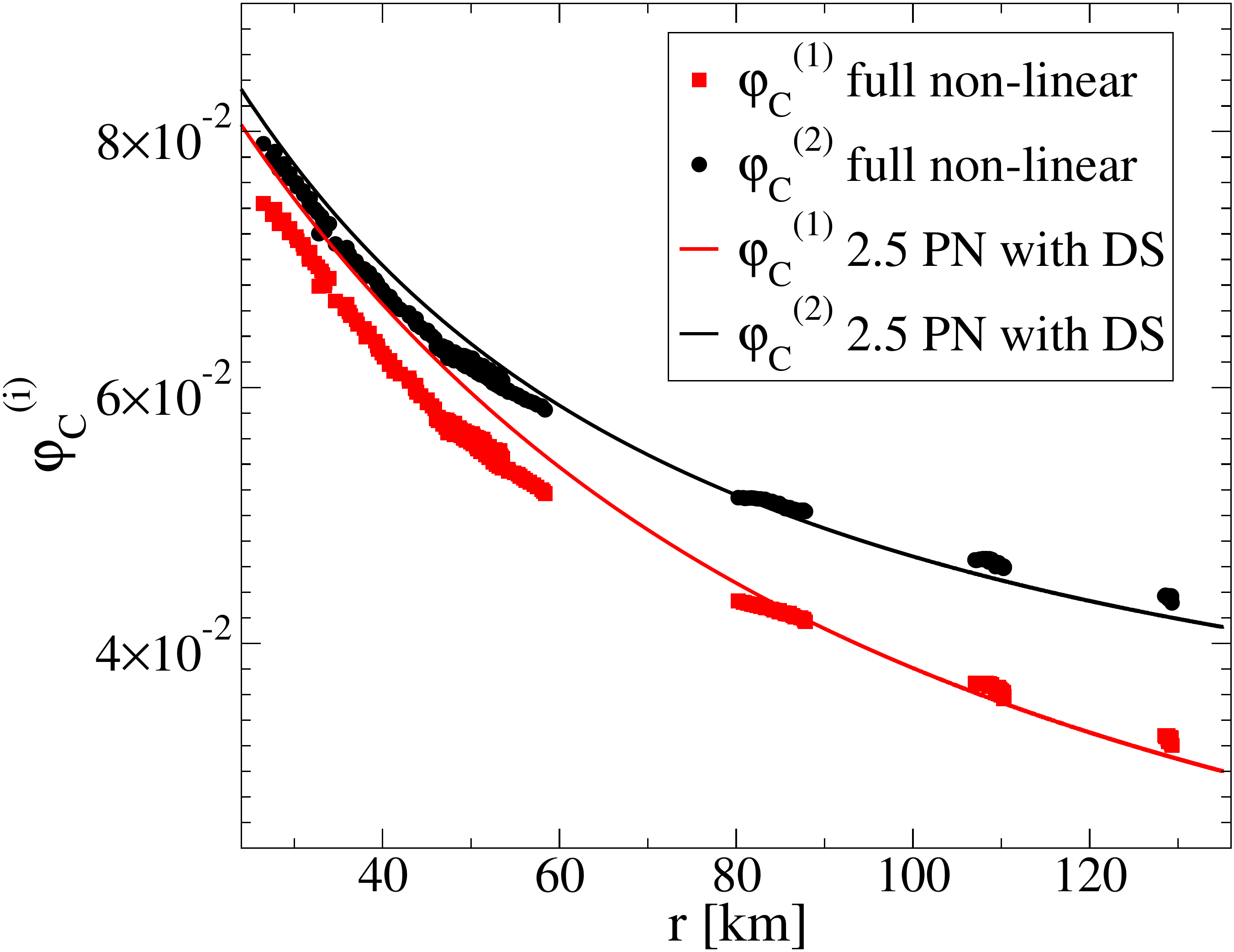}\\
\caption{\footnotesize Central values of the scalar field $\varphi_C^{(i)}$ 
as a function of the binary separation for the unequal mass case
$m_1=1.64 M_{\odot}$, $m_2=1.74 M_{\odot}$ with ${\tilde \beta} = -4.5$ and $\varphi_{0} G^{1/2}=10^{-5}$.
Results obtained with full non-linear simulations are represented by circles and squares, while those obtained with 2.5PN evolutions accounting for DS are represented with solid lines.
[Note that the vertical scale is linear, not logarithmic as in Fig.~\ref{fig:phic_r_mass15_15}.]
\label{fig:phic_r_mass16_17}}
\end{figure}

For another illustration of the behavior displayed by the system -- and the way our
model successfully captures it -- Fig.~\ref{fig:phic_r_mass16_17} shows the results
for an  unequal mass binary with masses $m_1=1.64 M_{\odot}$ and $m_2=1.74 M_{\odot}$. This system
was studied until merger from an initial separation of $70$ km with a fully non-linear simulation in Ref.~\cite{ourselves}. Here, we also show three additional  
larger separations, at which the system is evolved
for about one orbit in order to extract the central value of the scalar field after the relaxation of the initial data.
Note that for this binary, one of the stars is compact enough to scalarize spontaneously
in isolation, which leads to a strong DS of the other star. The agreement obtained
with the model that we introduced in this paper provides evidence that DS is able to correctly, and efficiently, capture the overall behavior
of the system.

With a validated model, we are then in a position to explore the physical parameter
space of NS binaries and examine their phenomenology in scalar-tensor theories. 
We adopt an asymptotic value for the scalar field
$\varphi_{0} G^{1/2}=10^{-5}$, and examine several cases varying the individual
masses of the binary's components as well as $\tilde \beta$.  We recall that based on the discussion
of the previous sections, the results are not expected to
be very sensitive to $\varphi_{0}$ (as long as one restricts attention to
viable values only), and that we consider $\tilde \beta \geq -4.5$.
To characterize the wavestrain produced by the systems under study, we consider the projection of the
gravitational wave on to the $l=2,m=2$ spin-weighted $s=-2$ spherical harmonic and normalize it with
respect to the observer's distance and total mass from the system,
\begin{equation}
h_{22} \equiv \frac{R}{M}<h_+ - i h_\times,^{-2}Y_{2,2}>  \, .
\end{equation}
The strain is calculated at leading order (i.e. using the standard
quadrupole formula) with the trajectories obtained by evolving
the 2.5PN equations of motion. We stress that the dipolar scalar mode
couples weakly to a gravitational-wave detector far from the source
(indeed, as showed in Ref.~\cite{ourselves}, the coupling to the detector vanishes as $\varphi_0\to0$), and  is
therefore not observable directly. However, the dipole channel
still carries energy and angular momentum away from the source, thus backreacting on the binary's trajectory and 
on its leading-order quadrupolar emission.

\subsection{Equal-mass binaries}

\begin{figure}
\centering
\includegraphics[height=7.0cm,width=8.5cm,angle=0]{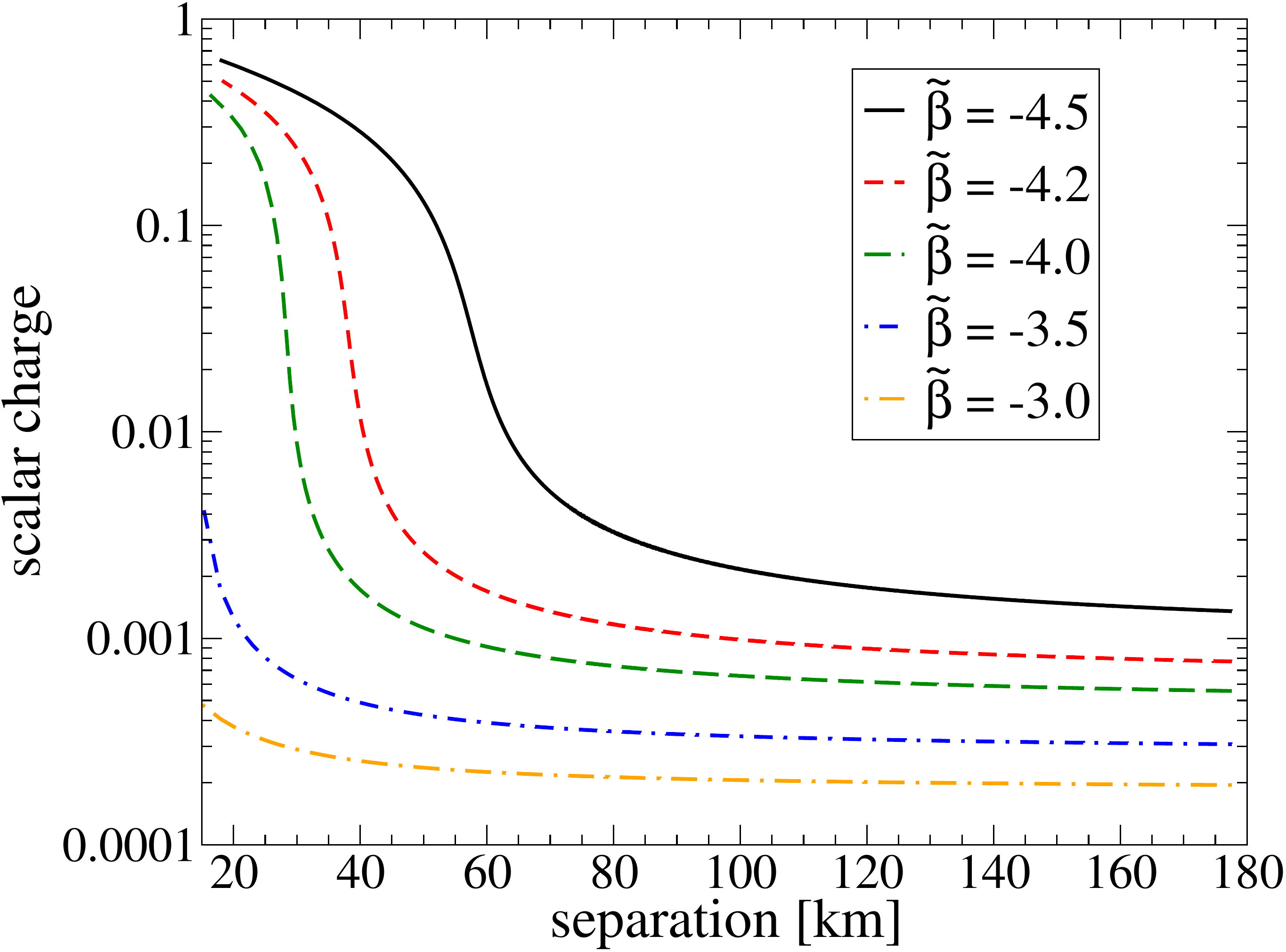}
\caption{\footnotesize Scalar charge of each star in an equal-mass binary with $m_1=m_2=1.51 M_{\odot}$, at
different separations and for different values of $\tilde \beta$. As the orbital separation decreases,
the scalar charges increase.
\label{fig:PNsims_}}
\end{figure}

We first concentrate on equal-mass binaries. Fig.~\ref{fig:PNsims_} illustrates the
value of the scalar charge for a binary with masses $m_1=m_2=1.51 M_{\odot}$ as a function of
separation, for
different values of  $\tilde \beta$. As discussed earlier, the scalar charge grows
due to DS as the orbit shrinks. For $\tilde \beta<-4$ the scalar
charge grows to $>0.1$ near the coalescence regime, and reaches
those values earlier as $\tilde \beta$ is reduced. For instance, the orbital frequency
at which $\alpha=0.1$ is reached is $\approx 1557$ rad/s for $\tilde \beta=-4.5$ and $\approx 4140$ rad/s for $\tilde \beta=-4$.

\begin{figure}
\centering
\includegraphics[height=7.0cm,width=8.5cm,angle=0]{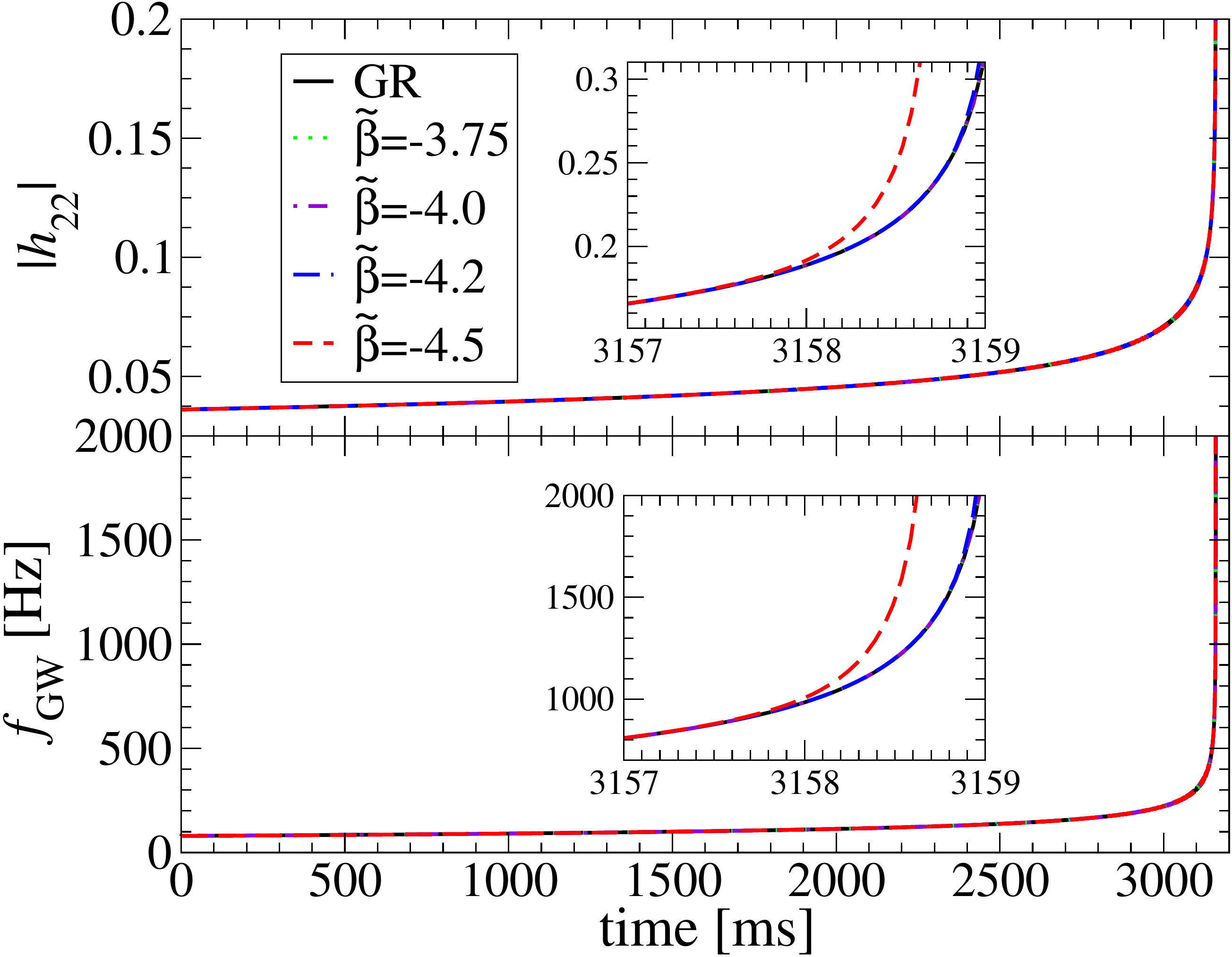}
\caption{\footnotesize Normalized gravitational-wave strain
and frequency for an equal mass binary (with $m_1=m_2=1.41 M_{\odot}$)
as a function of time, for different values of $\tilde \beta$. All cases display essentially the same quantitative behavior, with small
departures from GR at the onset of the plunge, as can be appreciated in the insets and in Fig.~\ref{fig:hf_m141_m141}.
\label{fig:equal141wavestime}}
\end{figure}

\begin{figure}
\centering
\includegraphics[height=7.0cm,width=8.5cm,angle=0]{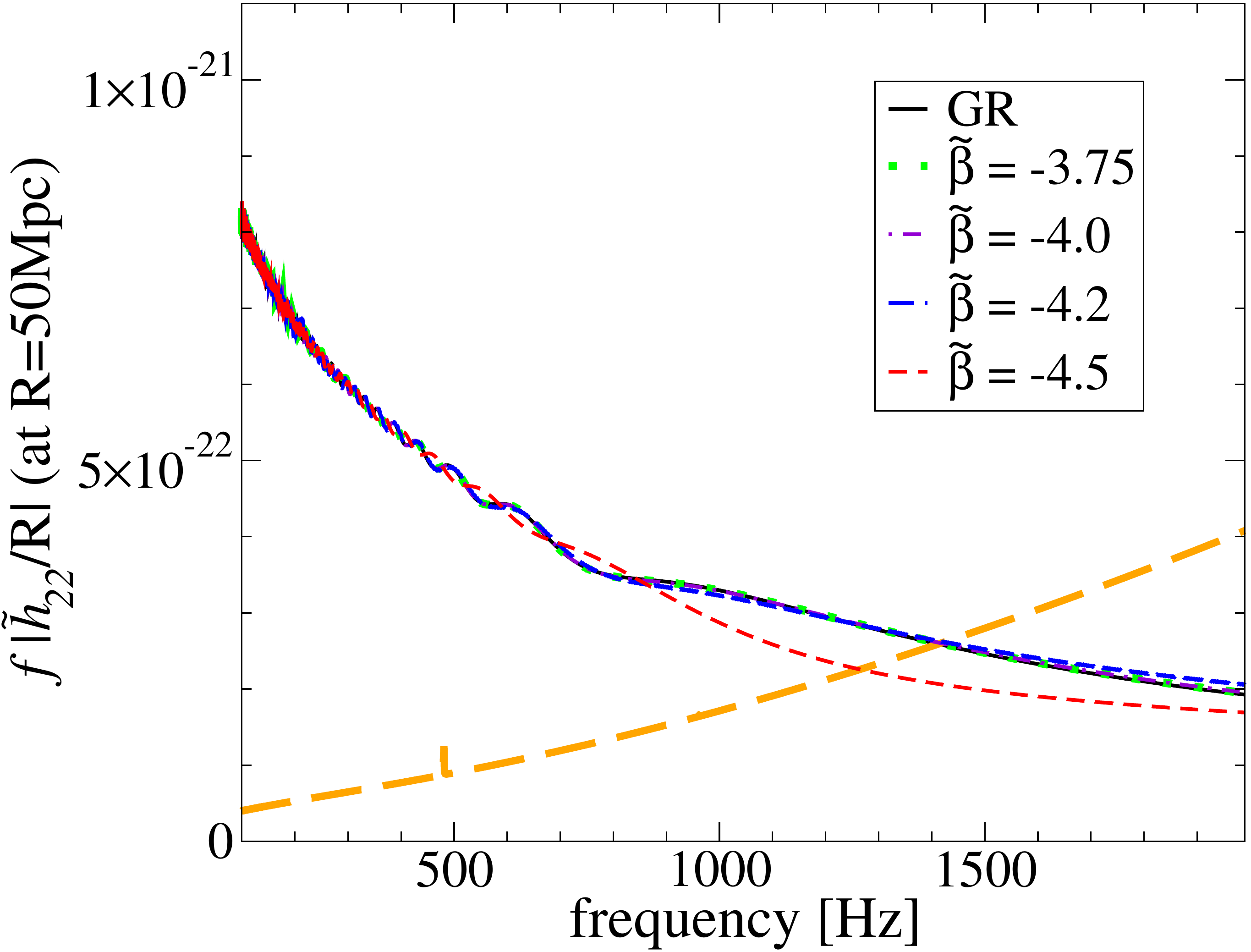}
\caption{\footnotesize Fourier spectrum of the $l=m=2$ gravitational-wave mode for a binary with $m_1=m_2=1.41 M_{\odot}$, 
for different values of ${\tilde \beta}$. The dashed, monotonic line shows $\sqrt{f  S_n(f)}$,
where $S_n{(f)}$ is the noise power spectrum of Advanced LIGO. Departures from GR become visible at $f \simeq 800$ Hz for $\tilde \beta = -4.5$.
\label{fig:hf_m141_m141}}
\end{figure}

As another example, the (renormalized) wavestrain and the frequency of the gravitational wave 
produced by a binary with
equal masses $m_1=m_2=1.41 M_{\odot}$ are represented in Fig.~\ref{fig:equal141wavestime},
for different values of $\tilde \beta$, together with the result expected in GR.
This case shows little difference from the GR behavior, for $\tilde \beta\gtrsim -4.2$. This is clearly illustrated in Fig.~\ref{fig:hf_m141_m141},
which plots the Fourier spectrum of the gravitational mode $l=m=2$ as measured at a distance
of $50$ Mpc. 
Only for the most extreme case $\tilde \beta=-4.5$ do differences arise
at high frequencies. For reference, the plot (as well as analog ones for the other cases) also 
includes the estimated noise power spectrum of Advanced LIGO~\cite{ligocurve}.

\begin{figure}
\centering
\includegraphics[height=7.0cm,width=8.5cm,angle=0]{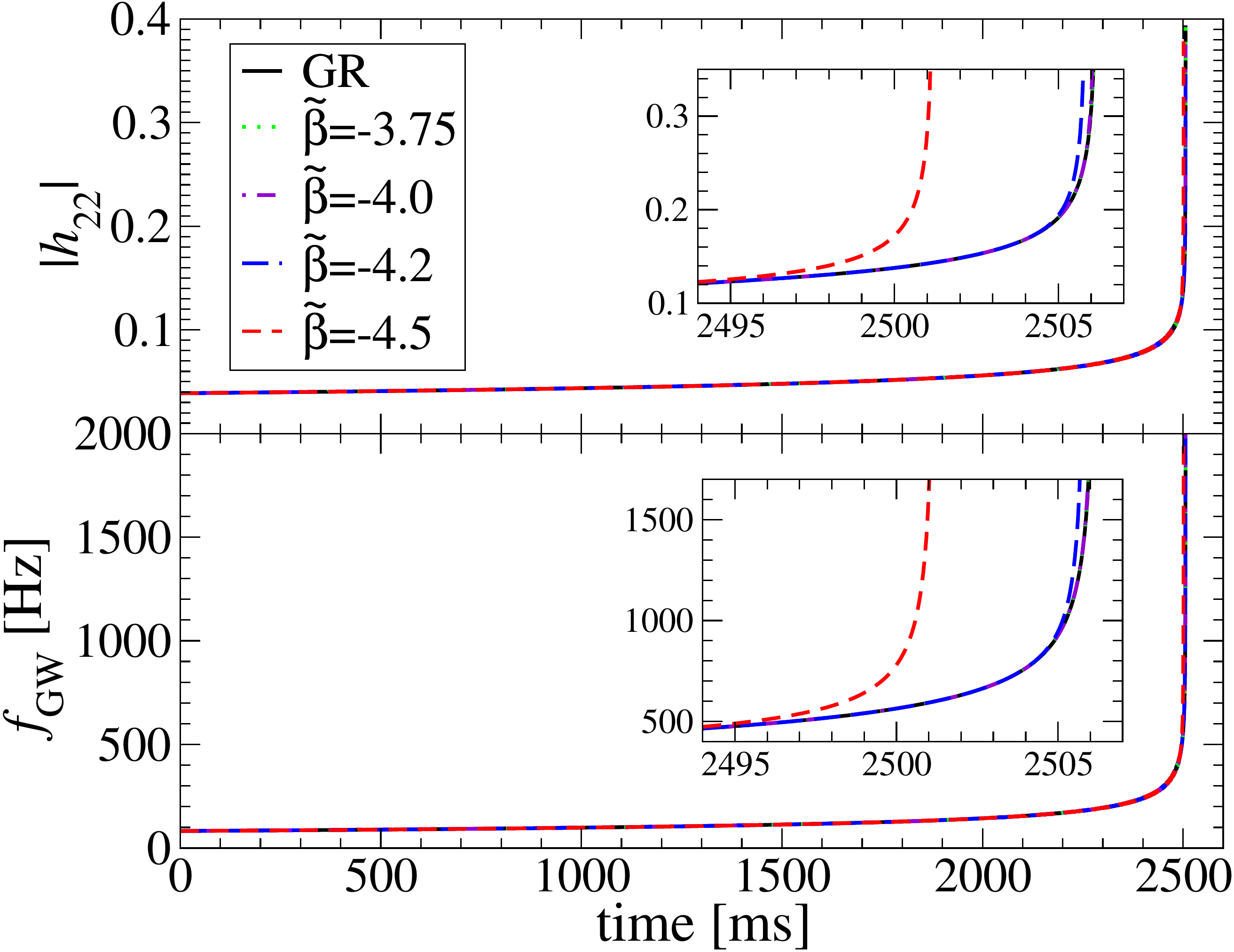}
\caption{\footnotesize Same as Fig.~\ref{fig:equal141wavestime} but for a binary with $m_1=m_2=1.51 M_{\odot}$. 
All cases display similar quantitative behavior, except
near merger, where  earlier plunges are triggered for $\tilde \beta = -4.5$, and in less severe manner
for $\tilde \beta = -4.2$. These departures from the GR expectation can be seen more distinctly in the insets and 
in Fig.~\ref{fig:hf_m151_m151}.
\label{fig:equal151wavestime}}
\end{figure}

\begin{figure}
\centering
\includegraphics[height=7.0cm,width=8.5cm,angle=0]{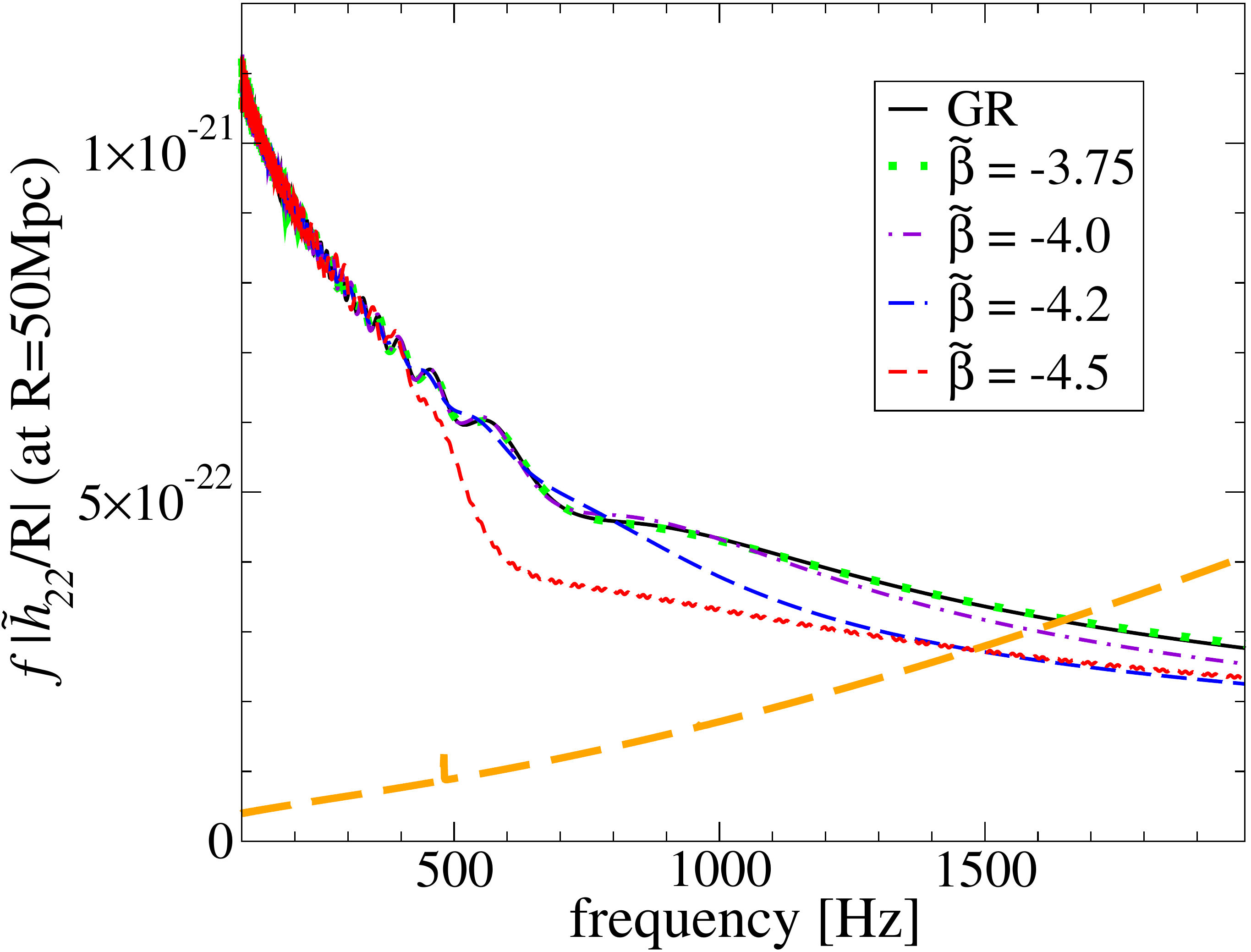}
\caption{\footnotesize Same as Fig.~\ref{fig:hf_m141_m141} but for a binary with $m_1=m_2=1.51 M_{\odot}$. 
Deviations from the expected GR behavior arise as early as $500$ Hz for $\tilde \beta =-4.5$, and $800$ Hz for $\tilde \beta =-4.2$.
\label{fig:hf_m151_m151}}
\end{figure}

As discussed, stronger scalarization effects arise when compactness increases. It is thus instructive
to examine the dynamics of more massive binaries (i.e. ones for which the effective compactness $\bar{C}$ is larger at  a given separation). 
Figs.~\ref{fig:equal151wavestime} and
\ref{fig:hf_m151_m151} show the same quantities as in the previous case, but for an equal-mass
binary with $m_1=m_2=1.51 M_{\odot}$. As can be seen, departures from the expected GR behavior
are more marked, showing differences already at $f\lesssim 700$ Hz for $\tilde \beta \lesssim -4.2$.

\begin{figure}
\centering
\includegraphics[height=7.0cm,width=8.5cm,angle=0]{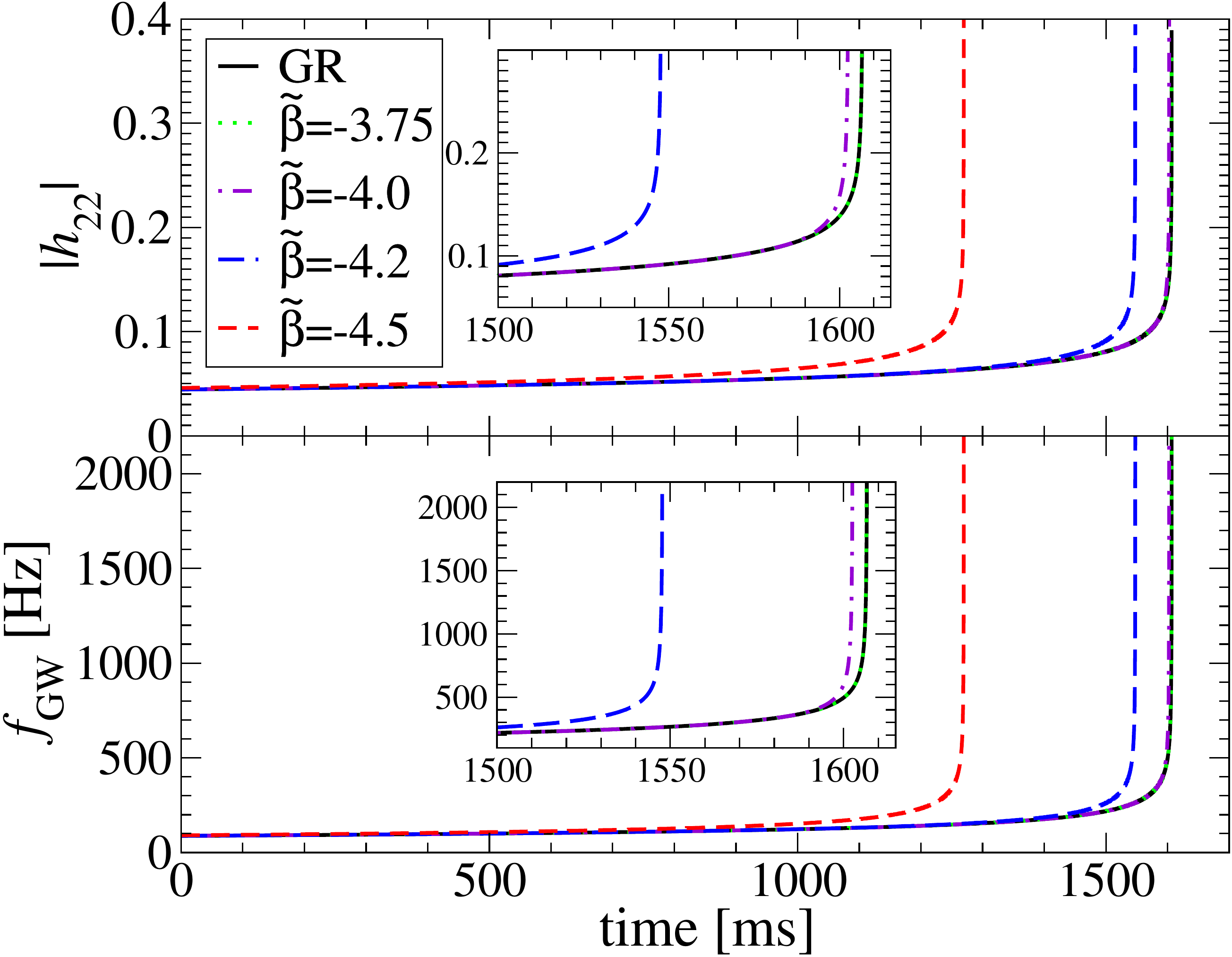}
\caption{\footnotesize Same as Fig.~\ref{fig:equal141wavestime} but for a binary with $m_1=m_2=1.74 M_{\odot}$. 
For this more massive binary,  plunges take place significantly earlier than in GR -- by about 600 ms --
for $\tilde \beta = -4.5$, because the stars spontaneously scalarize in isolation. For the cases with 
$\tilde \beta = -4.2, -4.$, DS induces an earlier plunge by about $200$ ms and  $10$ ms respectively. The consequences
of this behavior for gravitational-wave detection are illustrated in Fig.~\ref{fig:hf_m174_m174}.
\label{fig:equal174wavestime}}
\end{figure}

\begin{figure}
\centering
\includegraphics[height=7.0cm,width=8.5cm,angle=0]{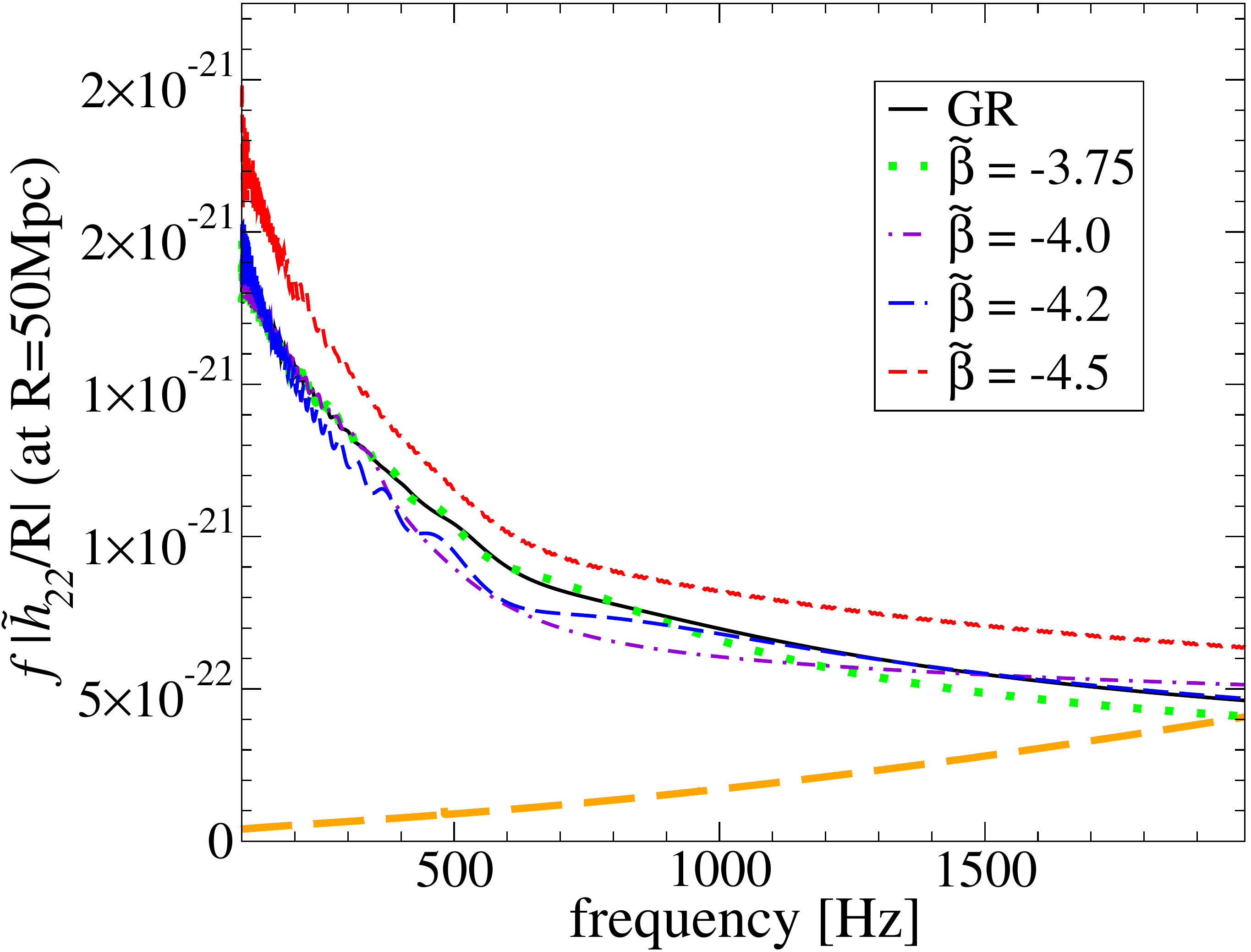}
\caption{\footnotesize Same as Fig.~\ref{fig:hf_m141_m141} but for a binary with
$m_1=m_2=1.74 M_{\odot}$. 
Distinct differences are evident already at $f\simeq 400$ Hz for $\tilde \beta \lesssim -4$, and throughout for $\tilde \beta = -4.5$.
This is because the binary's components spontaneously scalarize in isolation.
\label{fig:hf_m174_m174}}
\end{figure}

For an even more massive binary, with $m_1=m_2=1.74 M_{\odot}$, noticeable departures arise at even lower frequencies, as can be seen in
Figs.~\ref{fig:equal174wavestime} and~\ref{fig:hf_m174_m174}.
Several features can be observed in these figures.
For $\tilde \beta=-4.5$ the stars scalarize in isolation. This means that departures from
the GR predictions are evident at all frequencies.  For larger values of $\tilde \beta$ -- for which spontaneous scalarization does
not take place in isolation -- one can see the effects of DS inducing departures from the 
GR behavior already at $f\simeq 300$ Hz for $\tilde \beta \lesssim -4$.

\subsection{Unequal mass binaries}
\begin{figure}
\centering
\includegraphics[height=7.0cm,width=8.5cm,angle=0]{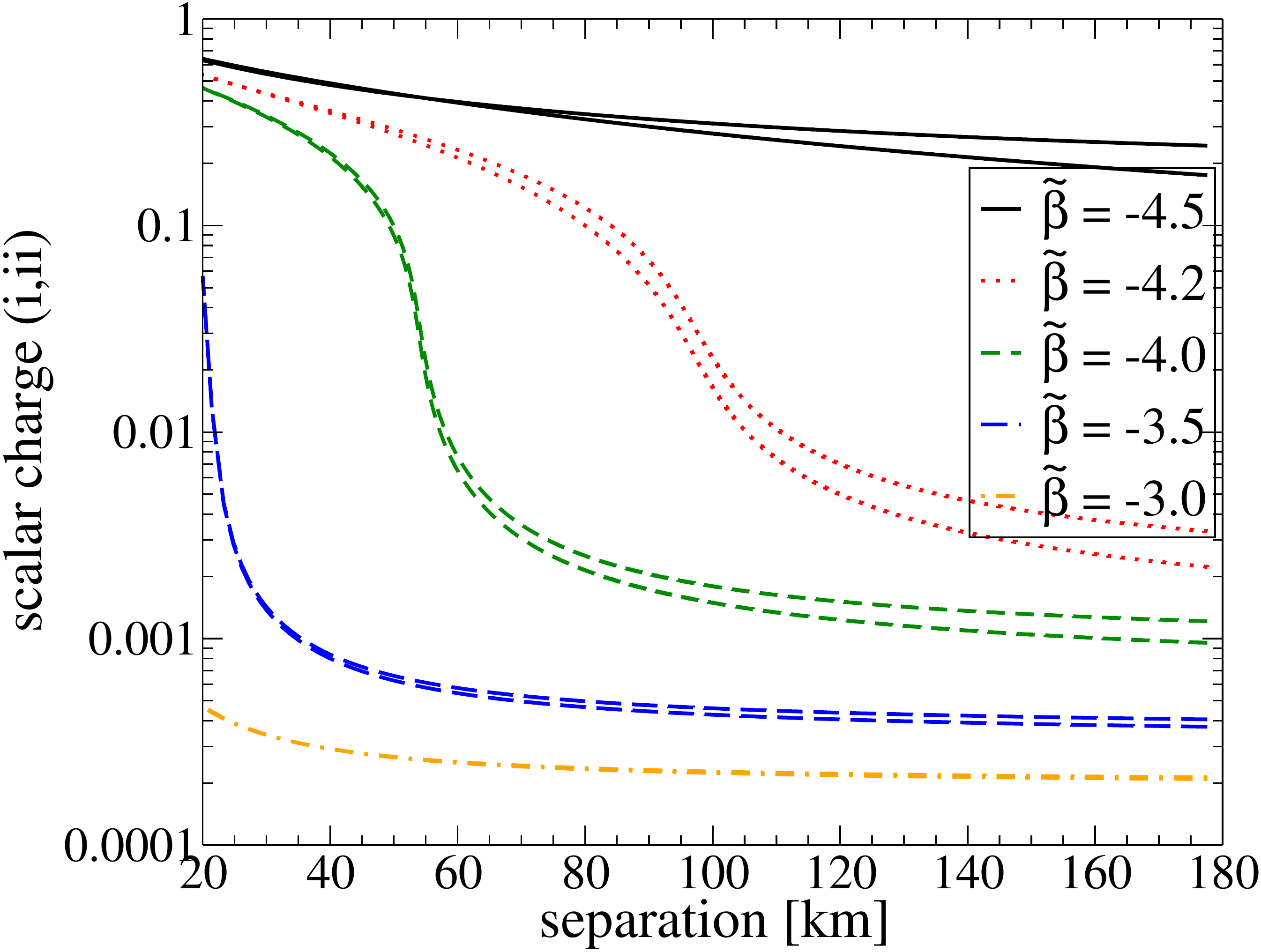}
\caption{\footnotesize Scalar charges for an unequal-mass binary with $m_1=1.74 M_{\odot}$ and $m_2=1.64 M_{\odot}$, as a function of
separation. The higher value of the scalar charge corresponds to the more massive star, as expected on the basis of its
 stronger gravitational field.
\label{fig:unequalcharges}}
\end{figure}

\begin{figure}
\centering
\includegraphics[height=7.0cm,width=8.5cm,angle=0]{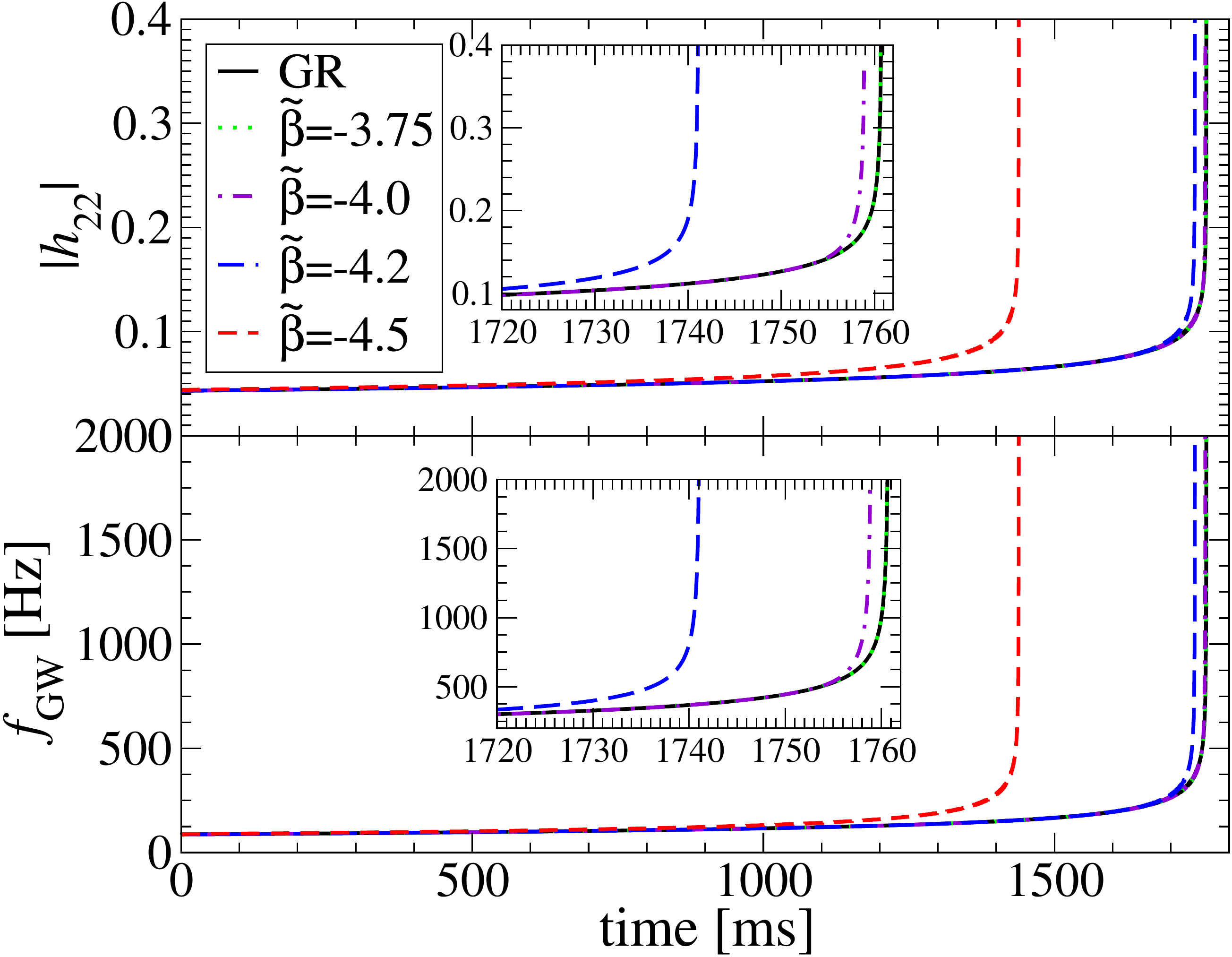}
\caption{\footnotesize Same as Fig.~\ref{fig:equal141wavestime} but for a binary with $m_1=1.64 M_{\odot}, m_2=1.74 M_{\odot}$. 
Again, $\tilde \beta = -4.5$ induces an earlier plunge than in GR -- by about $550$ ms --
as in this case one star is scalarized from the onset and both IS and DS play a strong role.
For $\tilde \beta = -4.2$, DS also induces an earlier plunge than in GR --by about $40$ ms.
\label{fig:164174vstime}}
\end{figure}

\begin{figure}
\centering
\includegraphics[height=7.0cm,width=8.5cm,angle=0]{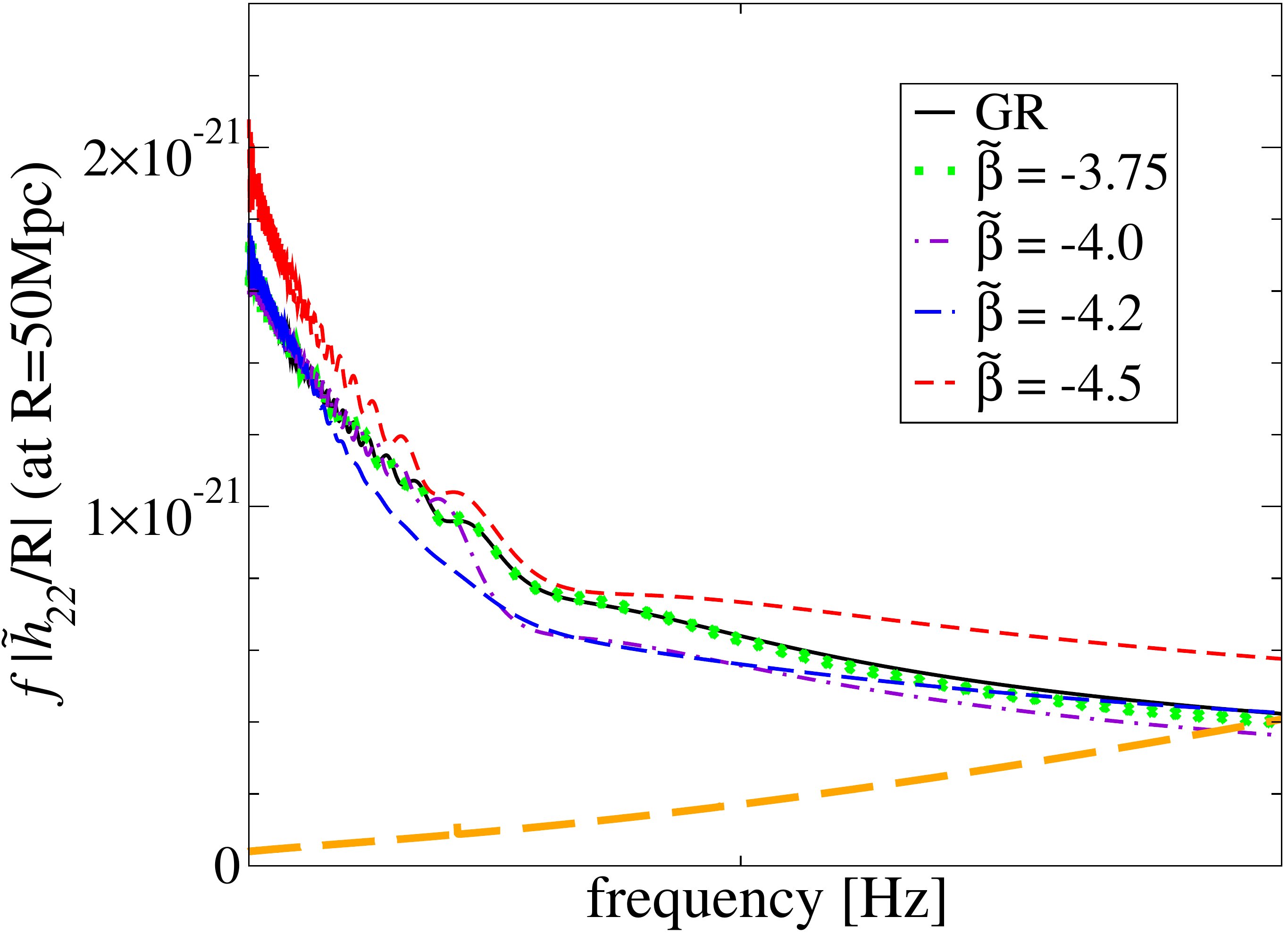}
\caption{\footnotesize Same as Fig.~\ref{fig:hf_m141_m141} but for a binary with $m_1=1.64 M_{\odot}$ and $m_2=1.74 M_{\odot}$.
Clear differences are evident already at $f\simeq 400$ Hz for $\tilde \beta \lesssim -4$, and throughout for $\tilde \beta = -4.5$, because
in this case the more massive star spontaneously scalarizes in isolation.
\label{fig:hf_m164_m174}}
\end{figure}

Currently known binary pulsars have unequal masses, thus it is important to consider the phenomenology
exhibited by unequal-mass binaries. Our model allows for a straightforward calculation
of these cases as well. For instance, Fig.~\ref{fig:unequalcharges} illustrates
the scalar charges for each of the stars of a binary with $m_1=1.74 M_{\odot}$ and $m_2=1.64 M_{\odot}$. As discussed, the scalar charges
grow as the separation decreases because of DS (as well as IS for the case
with $\tilde \beta=-4.5$ since, as noted, the more massive star in this binary scalarizes in isolation). The charges
become of order $0.1$ at considerably early times in the dynamics for $\tilde \beta < -4$.
For this case, the behavior of $|h_{22}|$ and that of the gravitational-wave
frequency as a function of time are presented in Fig.~\ref{fig:164174vstime}. Strong departures from the GR
expectation are evident for the case $\tilde \beta = -4.5$, and seemingly more
subtle ones for $\tilde \beta = -4.2$ and above. A more clear appreciation of these differences and their
impact on gravitational-wave detection can be obtained
by examining the Fourier spectrum of the gravitational signal.  Fig.~\ref{fig:hf_m164_m174} 
shows the resulting behavior as measured by an observer at $50$ Mpc. Clear differences arise already at $f_{GW} \simeq  400$ Hz and $600$ Hz for
$\tilde \beta = -4.2$ and $-4.0$ respectively. Also, the case $\tilde \beta = -4.5$ shows
departures throughout as the more massive star is scalarized from the onset.

\begin{figure}
\centering
\includegraphics[height=7.0cm,width=8.5cm,angle=0]{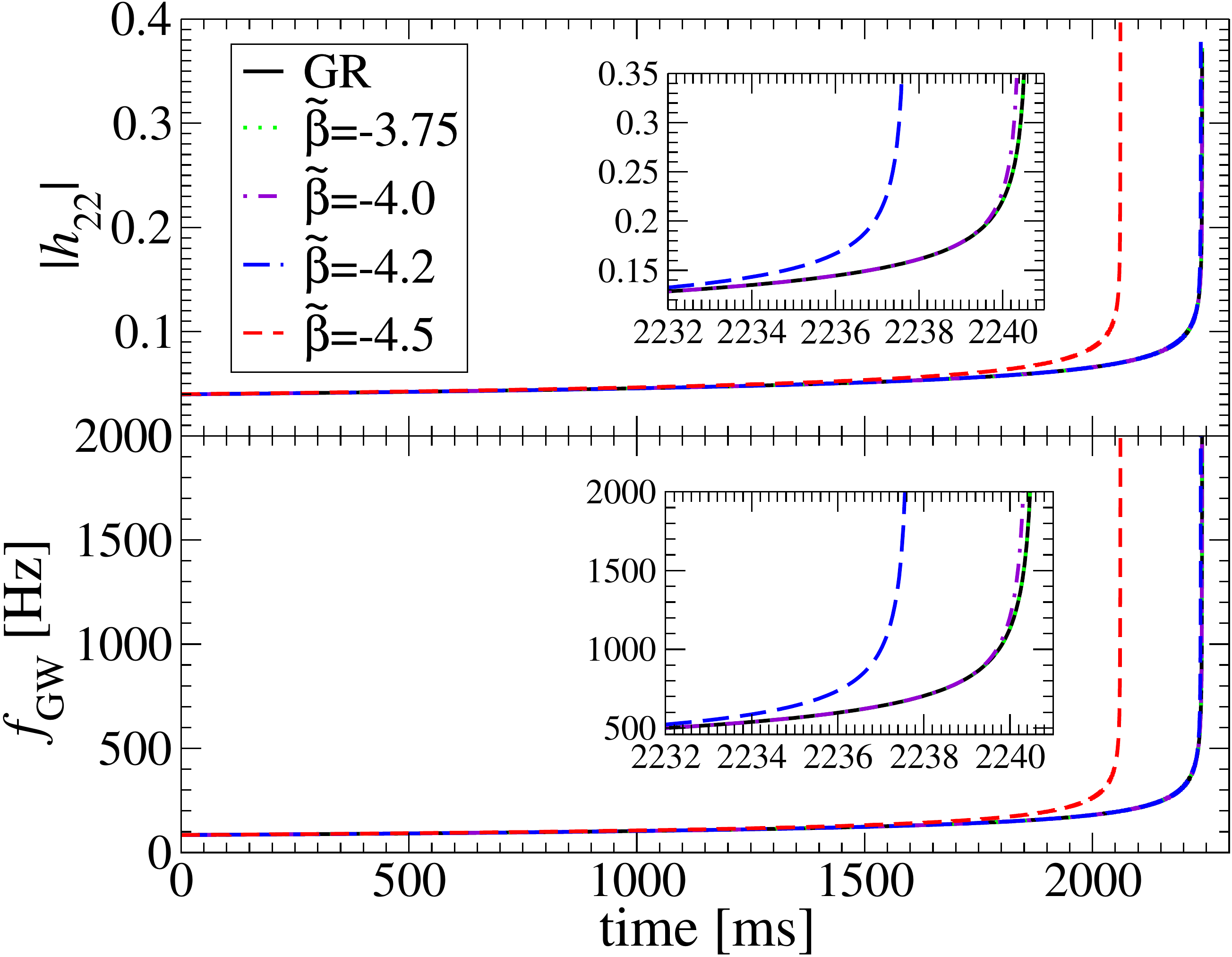}
\caption{\footnotesize Same as Fig.~\ref{fig:equal141wavestime} but for a binary with
$m_1=1.41 M_{\odot}, m_2=1.74 M_{\odot}$. Increasingly earlier plunges are induced as $\tilde \beta$ decreases.
\label{fig:141174vsfreq}}
\end{figure}

\begin{figure}
\centering
\includegraphics[height=7.0cm,width=8.5cm,angle=0]{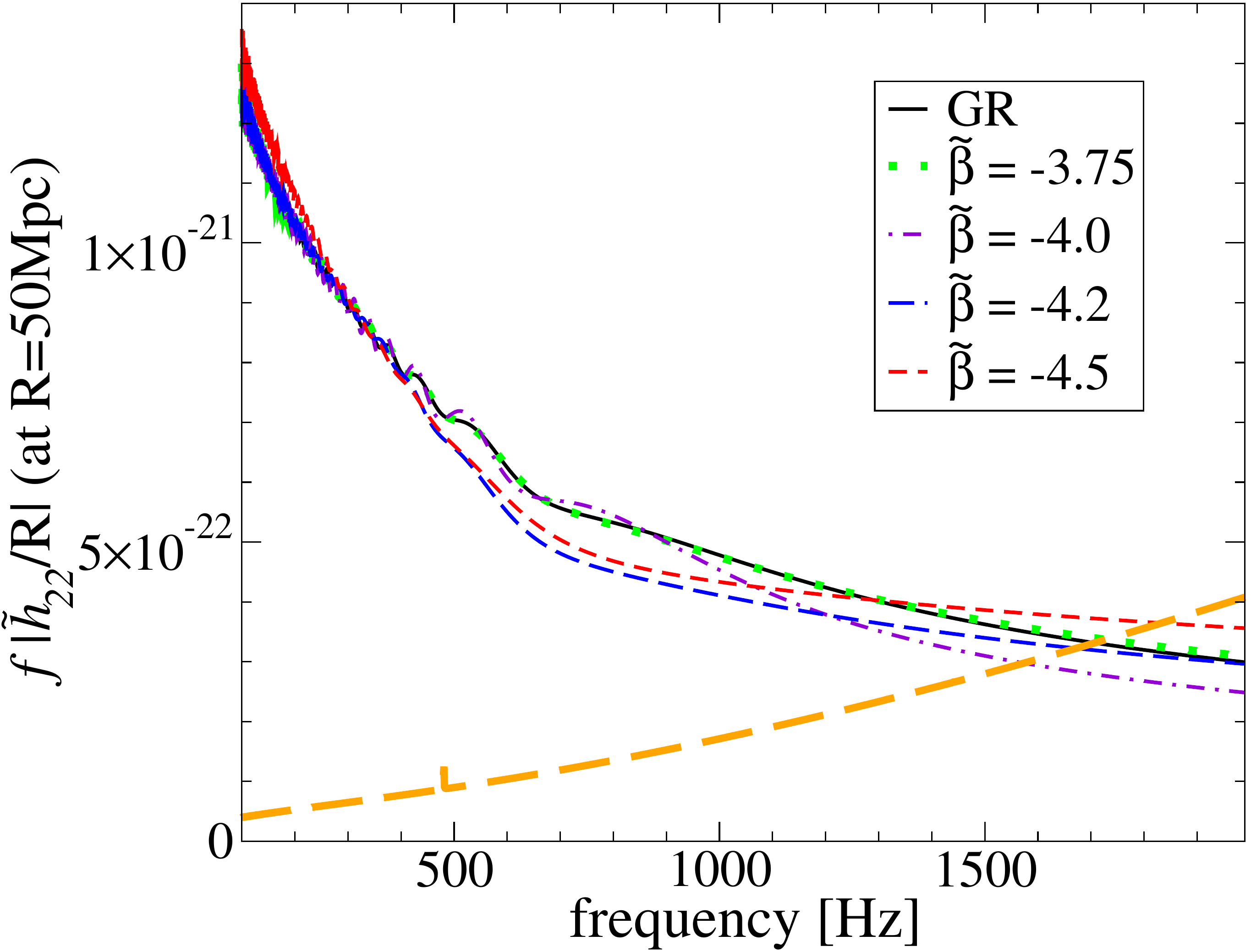}
\caption{\footnotesize 
Same as Fig.~\ref{fig:hf_m141_m141} but for a binary with
$m_1=1.41 M_{\odot}$,$m_2=1.74 M_{\odot}$. Departures from the GR expectation begin
at  $600$ Hz for $\tilde \beta = -4.2$ and $800$Hz for  $\tilde \beta = -4.$. The case $\tilde \beta = -4.5$ presents
differences throughout, as the more massive star spontaneously scalarizes in isolation.
\label{fig:hf_m141_m174}}
\end{figure}

The results for an unequal mass binary with $m_1=1.41 M_{\odot}$ and $m_2=1.74 M_{\odot}$
are presented in Figs.~\ref{fig:141174vsfreq} and~\ref{fig:hf_m141_m174}. This binary
experiences earlier plunges as $\tilde \beta$ is decreased. In fact, for $\tilde \beta = -4.5$,
the plunge sets in about $500$ ms earlier than in GR. The different dynamical behaviors are
clearly distinguished in Fig.~\ref{fig:hf_m141_m174}, which shows significant differences from GR for $\tilde \beta \lesssim -4.0$.

\subsection{Eccentric binaries}

Recently, it has been proposed~\cite{O'Leary:2008xt,Lee:2009ca,Samsing:2013kua,
Antonini:2013,Seto:2013} that eccentric binaries, formed via dynamical
capture and binary-single scatterings,
might  contribute to the population of sources measurable by upcoming gravitational-wave detectors.
These eccentric binaries, while not expected
to constitute a large fraction of the measurable sources for Advanced LIGO/VIRGO, display nevertheless interesting phenomenology,
which allows one to extract key properties of the system. For instance, they are believed to produce massive disks around BHs, 
produce a zoom-whirl behavior that affects gravitational-wave emission, etc. 
It is thus important to examine what additional phenomenology might arise within scalar-tensor theories. In particular,
these binaries could give rise to a {\em transient dynamical scalarization}. At an intuitive level,
this would be caused by the increase in the effective compactness of the system as the objects approach, which would trigger
DS, but then when the objects move apart, de-scalarization should occur as the effective compactness decreases.
We illustrate that this behavior indeed takes place with some representative cases already studied in full GR~\cite{Gold:2011df,East:2012ww}.

\begin{figure}
\centering
\includegraphics[height=6.0cm,width=8.5cm,angle=0]{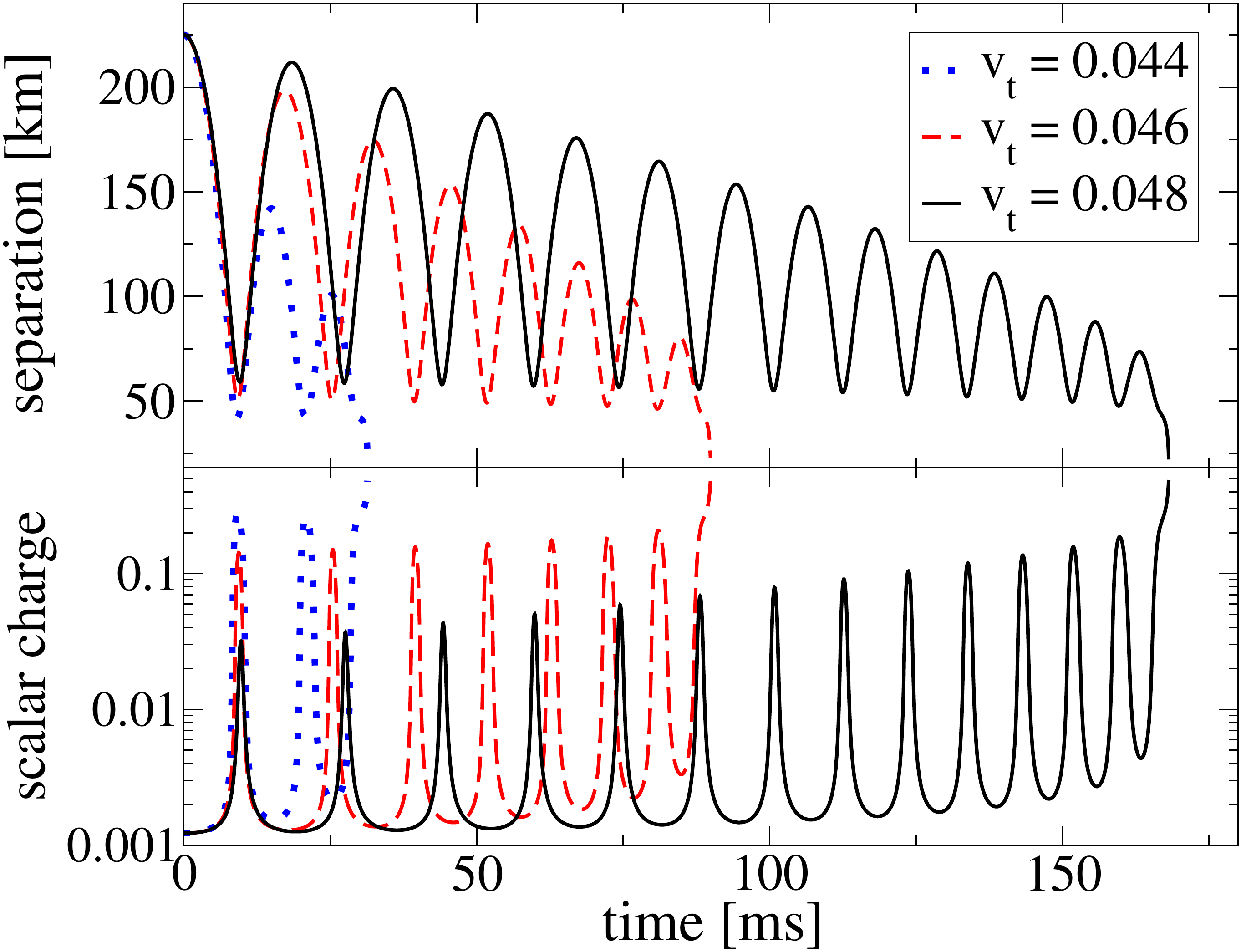}
\caption{\footnotesize Binary separation (top) and scalar charge (bottom) for an 
eccentric equal-mass binary $m_1=m_2=1.51 M_{\odot}$ with $\tilde \beta = -4.5$ and 
$\varphi_0 G^{1/2} = 10^{-5}$, for different values of the initial
tangential velocity $v_t$ (which determines the initial eccentricity). 
The evolution of the scalar charge mirrors the separation between the stars.
\label{fig:tbds_equal151_ralphatime}}
\end{figure}

Figure~\ref{fig:tbds_equal151_ralphatime} shows the separation (top panel) and the
scalar charge (bottom panel) for an equal-mass binary $m_1=m_2=1.51 M_{\odot}$
with $\tilde \beta = -4.5$ and $\varphi_0 G^{1/2} = 10^{-5}$, varying the initial eccentricity
(i.e., varying the initial tangential velocity $v_t$). In an equal-mass binary, dipolar radiation
is not produced, so the behavior observed is driven by the modified gravitational attraction between
the stars, which affects (through the dynamics) the emission of quadrupolar gravitational waves.
The eccentric behavior of the system is evident from the  oscillations in the separation. Such
oscillations induce a scalar charge, whose behavior is naturally tied to the separation. As expected, the scalar
charge increases as the separation decreases, achieving a maximum near each periastron. As the
separation increases the scalar charge is reduced, with a minimum near apastron.

\begin{figure}
\centering
\includegraphics[height=7.0cm,width=8.5cm,angle=0]{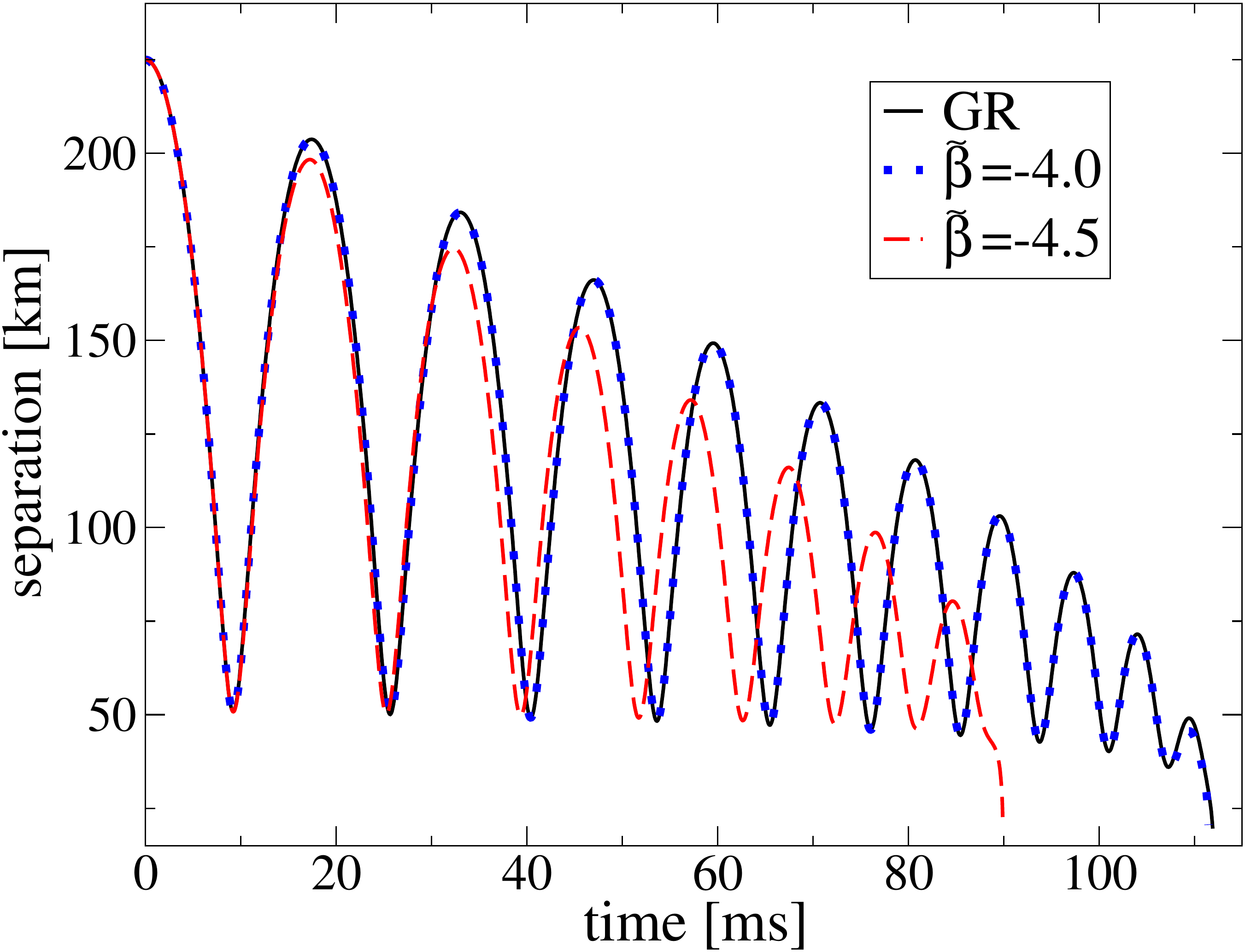}
\caption{\footnotesize Separation as a function of time for
an eccentric equal-mass binary (with $m_1=m_2=1.51 M_{\odot}$), for
$\varphi_0 G^{1/2} = 10^{-5}$ and an initial tangential velocity $v_t=0.046$, and
for different values of $\tilde \beta$. The 
conservative and dissipative gravitational effects due to DS help reduce the eccentricity of the system.
\label{fig:tbds_equal151_rtime}}
\end{figure}

\begin{figure}
\centering
\includegraphics[height=7.0cm,width=8.5cm,angle=0]{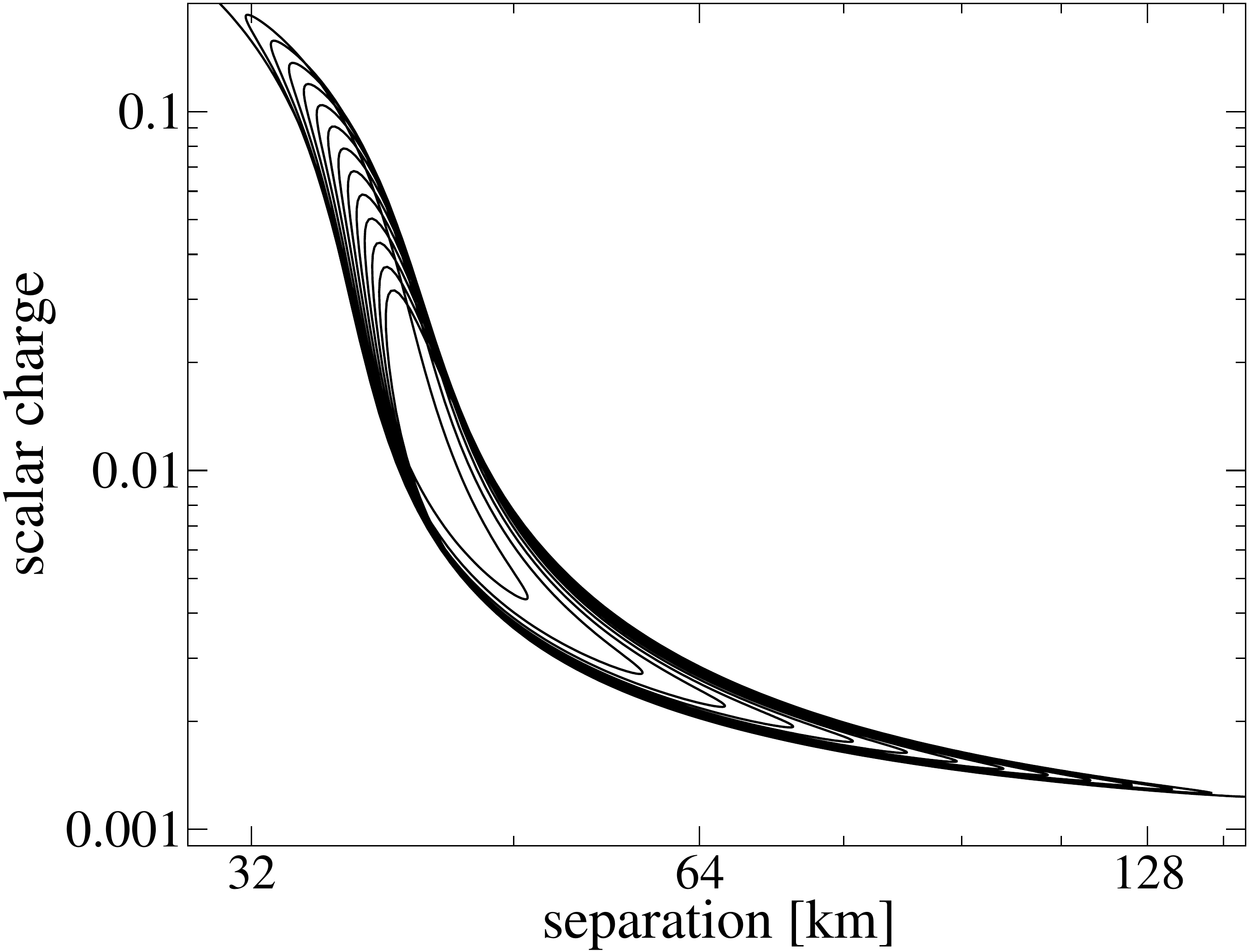}
\caption{\footnotesize Scalar charge as a function of separation
for an eccentric equal-mass binary with $m_1=m_2=1.51 M_{\odot}$,
$\varphi_0 G^{1/2} = 10^{-5}$, $\tilde \beta = -4.5$ and $v_t=0.046$. 
\label{fig:tbds_equal151_alphar}}
\end{figure}

Naturally this behavior has a strong impact on the dynamics of the system, as it induces
stronger gravitational attraction and enhanced radiation emission,  when compared to GR, at close separations.
This helps reduce the binary's eccentricity, as illustrated in Fig.~\ref{fig:tbds_equal151_rtime}, which
shows the separation as function of time for a given value of the initial eccentricity (corresponding
to $v_t=0.046$), and for two different values of $\tilde \beta$ (as well as for GR). For the smaller $\tilde \beta$,
eccentricity is reduced faster and an earlier plunge takes places. This transient DS
can be appreciated clearly by examining the value of the scalar charge as a function of separation.
For instance, Fig.~\ref{fig:tbds_equal151_alphar} illustrates the results for the case $\tilde \beta=-4.5$.
Note that the scalar charge behavior is not symmetric as the binary's trajectory is affected by the dissipative backreaction of gravitational
radiation.

\section{Final Comments}

We have presented a model that fully captures the behavior of binary NSs in scalar-tensor theories.
From a formal perspective, this model allows to naturally combine 
the different channels under which the system can scalarize: spontaneous, induced and dynamical
scalarizations are naturally encoded in our model. Our approach thus explains the connection
of these processes to the possible behavior of binary NSs in scalar-tensor theories. (The extension of this model to
other binaries is straightforward and the subject of ongoing work.) Furthermore, the insight provided by this model
allows for understanding general scenarios where dynamical scalarization might take place and how it would manifest. As an illustration
of this deeper understanding, we have presented the possibility 
of scalarization/de-scalarization phenomena that can take place in eccentric binaries.

From a practical point of view, this model allows for an efficient generation of templates of waveforms
within scalar-tensor theories, to explore possible degeneracies and examine the extent to which existing data analysis
techniques can be exploited to test for theories other than GR. Such studies will be presented elsewhere.
Beyond gravitational wave motivated applications, our model also stresses the importance of considering
the role of interactions in dynamical, non-linear gravitational systems. Scalar tensor theories are ubiquitous
in cosmological scenarios, and an extension of the ideas presented here may help highlight whether screening and related
mechanisms may be influenced/modified by dynamical interactions.
\\

\begin{acknowledgements}
We thank E. Poisson, N. Yunes, C. Hanna and K. Cannon
for comments and discussions as well as our collaborators
M. Anderson, E. Hirschmann, S.L. Liebling  and D. Neilsen with whom we have developed
the basic computational infrastructure employed in part of this work. 
We are especially grateful to Gilles Esposito-Farese for noting the subtlety 
explained in footnote 2.
We acknowledge support from
the European Union's Seventh Framework Programme (FP7/PEOPLE-2011-CIG)
through the  Marie Curie Career Integration Grant GALFORMBHS PCIG11-GA-2012-321608
(to E.B.); the Jeffrey L.~Bishop Fellowship (to C.P.)
and NSERC through a Discovery grant (to L.L.). We also acknowledge hospitality from
the Perimeter Institute (E.B.), 
where part of this work was carried out.
Computations were performed on the gpc supercomputer at the SciNet HPC Consortium and the HPC/titan at Perimeter Institute.
SciNet is funded by: the Canada Foundation for Innovation under the auspices of
Compute Canada; the Government of Ontario; Ontario Research Fund - Research Excellence;
and the University of Toronto.
Research at Perimeter Institute is supported through Industry Canada
and by the Province of Ontario through the Ministry of Research and Innovation.
\end{acknowledgements}

%
%

\appendix
\section{Convergence of the scalar field in DS}
\label{app:convergence}

\renewcommand{\theequation}{A-\arabic{equation}}
\setcounter{equation}{0}

The dynamics of the scalar fields is governed by the 
relation for the background field
\begin{equation}\label{proof_phiB}
  \varphi_{B}^{(i)} = \varphi_{0} 
+ \frac{\varphi_1^{(j)}}{r}
~~~~~~~~~i \neq j  ~~,~~ 
\end{equation}
with $\varphi_1^{(j)} = \varphi_1^{(j)} (\varphi_{B}^{(j)})$.
The convergence of the method can be assessed by constructing
the solution iteratively. At the step $0$ the background field
is just $^0 \varphi_B^{(i)} = \varphi_0$, while the solution
at the following order is obtained by substituting the current
solution in equations~(\ref{proof_phiB}). At the  $n+1$ step, 
the recursive relation is just
\begin{equation}\label{proof_phiB_stepn}
  ^{(n+1)}\varphi_{B}^{(i)} = \varphi_{0} 
+ \frac{^{(n)}\varphi_1^{(j)}}{r} ~~,~
\end{equation}

Subtracting this equation between steps $n+1$ and $n$, 
one obtains,
\begin{equation}\label{proof_phiB_substraction}
  \delta^{n+1}\varphi_{B}^{(i)} \equiv ^{(n+1)}\varphi_{B}^{(i)}
 - ^{(n)}\varphi_{B}^{(i)} = 
   \frac{^{(n)}\varphi_1^{(j)} - ^{(n-1)}\varphi_1^{(j)} }{r} 
\end{equation}

Let us now Taylor expand the second term in the right-hand-side of
the equations as,
\begin{equation}
 ^{(n-1)}\varphi_1^{(i)} \approx ^{(n)}\varphi_1^{(i)} -
  \frac{\partial \varphi_1^{(i)}}{\partial \varphi_B^{(i)}}
  \delta^{n}\varphi_{B}^{(i)} \, .
\end{equation}
Equations (\ref{proof_phiB_substraction}) can then be re-expressed in matrix form,
\begin{equation}
\left( \begin{array}{c}
\delta^{n+1}\varphi_{B}^{(1)} \\
\delta^{n+1}\varphi_{B}^{(2)} \end{array} \right) =
 \left( \begin{array}{cc}
0 & \frac{1}{r} \frac{\partial \varphi_1^{(2)}}{\partial \varphi_B^{(2)}}  \\
\frac{1}{r} \frac{\partial \varphi_1^{(1)}}{\partial \varphi_B^{(1)}} & 0 \end{array} \right)
\left( \begin{array}{c}
\delta^{n}\varphi_{B}^{(1)} \\
\delta^{n}\varphi_{B}^{(2)} \end{array} \right)
\end{equation}

A necessary condition for the iterative method to converge, which ensures
the difference between solutions at subsequent iterations decreases,
is that the (magnitude of) eigenvalues of the metric be bounded by 1. Thus,
$\delta^{n+1}\varphi_{B}^{(i)} < \delta^{n}\varphi_{B}^{(i)}$, if the following condition is satisfied
\begin{equation}\label{stability_solution_bis}
   \frac{1}{r} \sqrt{ \frac{\partial {\varphi}_1^{(1)}}{\partial \varphi_B^{(1)}}
                      \frac{\partial {\varphi}_1^{(2)}}{\partial \varphi_B^{(2)}}
                     } \le 1  .
\end{equation}


\end{document}